\documentclass[12pt,preprint]{aastex}

\slugcomment{PASP In Press}

\shorttitle{WFPC2 CTE Characterization}
\shortauthors{Dolphin}

\begin{document}


\title{A Revised Characterization of the WFPC2 CTE Loss}


\author{Andrew E. Dolphin}
\affil{Raytheon Company, Tucson, AZ 85756}
\email{adolphin@raytheon.com}

\begin{abstract}
Charge-transfer loss on the Wide Field Planetary Camera 2 (WFPC2) onboard
the Hubble Space Telescope is a primary source of uncertainty
in stellar photometry obtained with this camera.
This effect, discovered shortly after the camera was installed,
has grown over time and can dim stars by several tenths
of a magnitude (or even more, in particularly bad cases).
The impact of CTE loss on WFPC2 stellar photometry was characterized by several
studies between 1998 and 2000, but has received diminished attention since
ACS became HST's primary imager.
After the failure of ACS in January 2007, WFPC2 once again became the
primary imaging instrument onboard HST, restoring the importance of ensuring
accurate CTE corrections.

This paper re-examines the CTE loss of WFPC2, with three significant
changes over previous studies.
First, the present study considers calibration data obtained through
2007, thus increasing the confidence in the reliability of the CTE
corrections when applied to recent observations.
Second, the change in CTE loss during readout is accounted for analytically.
Finally, a reanalysis of the CTE dependencies on counts, background, and
observation date was made.  The
resulting correction is significantly more accurate than that provided in the
WFPC2 Instrument Handbook (Dolphin 2002 and updates through 2004),
resulting in photometry that can be enhanced by over 5\% in certain
circumstances.
\end{abstract}



\section{Introduction}

Shortly after its installation onboard the Hubble Space Telescope in 1994,
the Wide Field Planetary Camera 2 (WFPC2) was observed to suffer from
charge-transfer loss, a phenomenon in which charge is removed from a star's
image during CCD
readout \citep{hol95a}.  The effect of this inefficiency is to reduce the
star's apparent brightness, especially those with large Y (parallel-read
direction) values on the
CCDs (and, to a lesser extent, those with large X values).
An initial solution to this was to cool the camera from $-76^\circ$C
to $-88^\circ$C to reduce the magnitude of the effect, and to apply a
correction that scaled from linearly from zero for stars with $Y=0$ to
a maximum of 0.04 magnitudes for stars with $Y=800$.  \citet{hol95a} also
noted that the CTE loss appeared to be a function of background level, and
recommended no CTE correction for images with backgrounds over $\sim 250$
electrons.

This dependency of CTE on background, as well as dependencies on star
brightness and observation date, were quantified
by \citet{whi97} and updated by \citet{whi98}, who found that the CTE loss
in the Y direction could be corrected by an equation with the form
\begin{displaymath}
\hbox{CTS}_{corr} = \hbox{CTS} \times [ 1 + f_1(\hbox{CTS}) \times f_2(\hbox{BG}) \times f_3(\hbox{DATE}) \times Y ],
\end{displaymath}
where CTS is the star's brightness on the read-out image; BG is the
background level; DATE is the date of observation; and $f_1$, $f_2$, and
$f_3$ are used to denote functions.  (Whitmore et al. used power
law functions for both $f_1$ and $f_2$, and a linear function for $f_3$.)

In addition to reports from the WFPC2 team,
characterizations of CTE loss in the refereed literature were made by
\citet{ste98}, \citet{whi99},
\citet{sah00}, and \citet[hereafter D00]{dol00a}.  Differences in the assumed functional
form of the solution and in the photometry techniques resulted in
somewhat different correction prescriptions.  However, D00
found reasonable agreement between the corrected magnitudes being produced
by each of the four.

Since that time, there have been several significant issues requiring continued
examination of the effects of CTE loss on WFPC2 stellar photometry.
Most notable, given the time dependence of CTE loss,
is the question of the applicability of published CTE corrections
to data being currently obtained.  Especially given the
increased utilization of WFPC2 during cycles 15 and 16, reinvestigation of
CTE issues can greatly increase confidence in the photometry being generated.

Other concerns are related to the functional form of the CTE corrections
themselves, due to limited ranges of star brightness and background level
of the calibration observations.
\citet{whi02} examined the validity of existing CTE correction equations
on stars significantly fainter than had been previously examined.  While it
was encouraging that the D00 corrections were valid when applied to stars as
faint as
100 electrons (thus, the corrections are accurate for the vast majority of
data used by
WFPC2 science programs), for yet fainter stars these corrections appeared to
overestimate
the CTE loss by $\sim 0.2$ magnitudes.

Likewise, there is the open question of the applicability of existing CTE
corrections to higher background levels (specifically, those typical of
deep science exposures).
Early work on WFPC2 CTE loss relied primarily on data taken with very low
background levels (mostly less than 10 electrons), which required
assumptions to be
made regarding the effect of background level on CTE loss.  However, since
2000, WFPC2 calibration programs have increasingly made use of the camera's
preflash capability, in which the CCDs are pre-exposed with a flat-field
lamp before the actual observation.  (A preflashed short exposure with bright
stars is frequently used to
simulate the CTE loss seen in much longer observations of fainter
stars.)  These data permit
a significantly more accurate measurement of the effects of CTE loss on
higher-background data.

This paper re-evaluates the CTE corrections in light of these three concerns,
using calibration data obtained through August 2007.  Section 2 describes the
data used in this study, section 3 contains preliminary analysis of the
functional forms required for accurate CTE corrections, section 4 contains
the CTE corrections themselves, and section 5 examines effects of the
new corrections.

\section{Observations}

As with D00 and \citet[hereafter D02]{dol02}, the data used for this study were
observations of the WFPC2 calibration field in
\object[NGC 5139]{$\omega$ Cen}.  (
To select the data, an HST archive search was made within
1.5 arcmin of $\alpha=13^h25^m37^s$ and $\delta=-47^\circ35'58''$.)
Only cold-camera ($-88^\circ$C) images
were considered.  A total of 1184 data sets were photometered:
142 in F439W, 300 in F555W, 92 in F675W, 573 in F814W, and 77 in the other
$BVRI$ filters, with the most recent observations taken in August 2007.

The data were reduced using
HSTphot\footnote{http://purcell.as.arizona.edu/hstphot/} \citep{dol00b},
following the recommended
reduction recipe in the HSTphot manual except as follows.  First, in the
initial sky solution, the
``abbreviated'' mode of \tt{getsky} \rm was used.  Second, \tt{hstphot}
\rm  was executed
with an iterated sky solution during the photometry process (option 512),
CTE corrections turned off and photometry calibrated according to \citet{hol95b}
(option 32), and aperture photometry (option 1).  The lone option of note is
the use of aperture photometry, which increases photometric uncertainties
but is more robust
(PSF-fitting photometry can induce artificial nonlinearities if the PSF is
not \textit{exactly} correct) and thus better suited for this analysis.

To avoid complications resulting from the WF4 anomaly \citep{bir05},
all WF4 observations from SM3B (March 2002) onwards were omitted from this
study.

For this analysis, the WFPC2 data were compared against ground-based $BVRI$
calibration data of \citet{wal94}.  To facilitate this comparison, the $BVRI$
magnitudes were transformed to WFPC2 instrumental magnitudes using the
color terms of \citet{hol95b}.

After excluding data that failed to meet data quality requirements,
55286 stellar measurements were matched to the ground-based standard
data.  Statistics of the brightness, background, and observation date
distributions are presented in Table \ref{tab_stars}.

\section{Analysis}

Prior to attempting a generic solution for CTE loss as a function of
star brightness, background level, and date of observation; it is worthwhile
to examine each of these variables in isolation.  In order to do this,
the data have been selected to constrain two of these variables to limited
ranges, allowing an examination of the third.  For example, to analyze
CTE loss dependence on date of observation, a
limited range of star brightness and background level was chosen.  The
data were then divided into groups based on observation date, and each
group of data was fit as follows:
\begin{equation}
m_{WFPC2} - m_{ground} = a + b \frac{x}{800} + c \frac {y}{800},\label{eq_linfit}
\end{equation}
where the left hand side is the magnitude difference between the WFPC2 and
the ground-based standard data.  The fit coefficients $a$,
$b$, and $c$ give position-independent, XCTE, and YCTE terms respectively,
and are examined in the subsequent sections.

\subsection{Position-independent Errors}

\begin{figure}
\plottwo{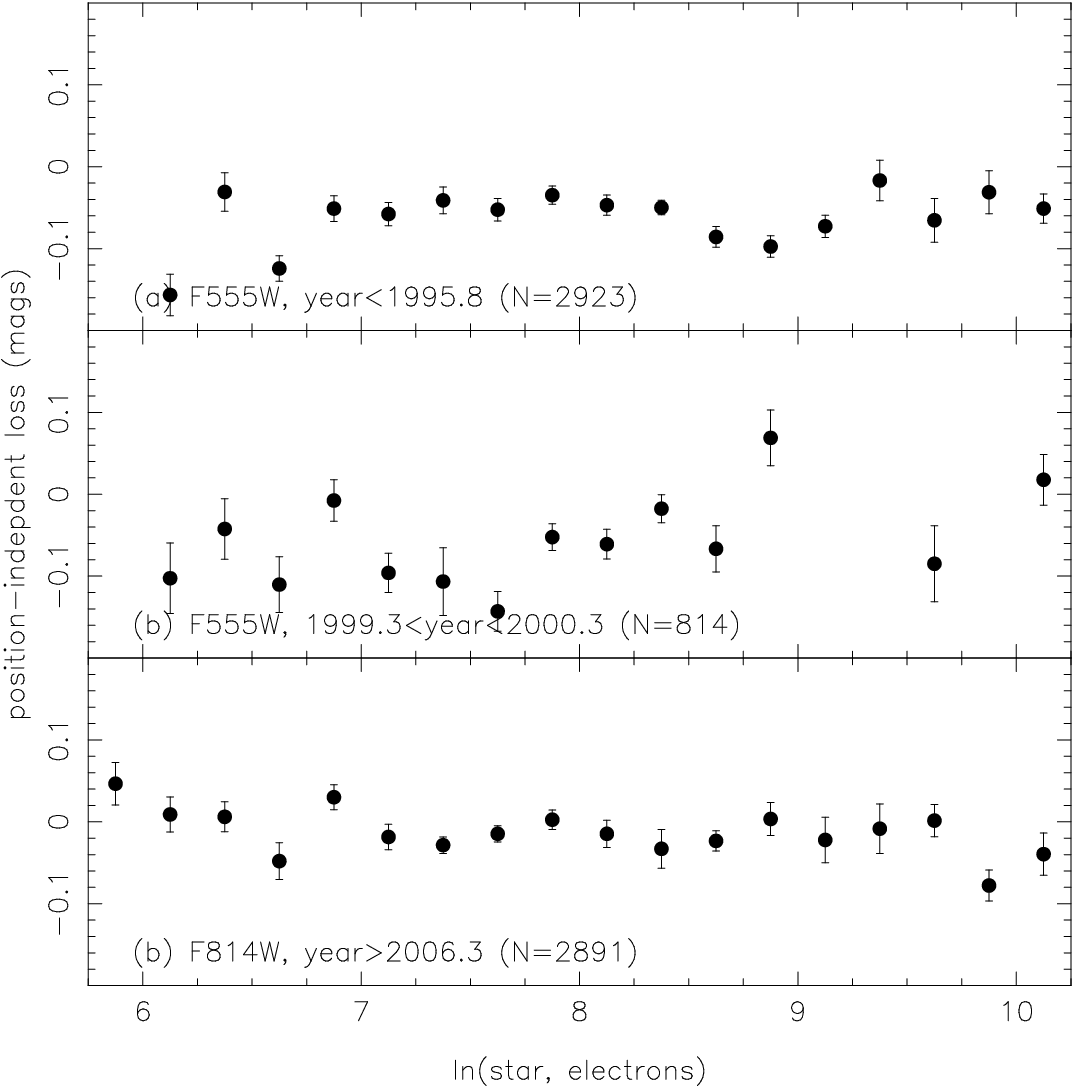}{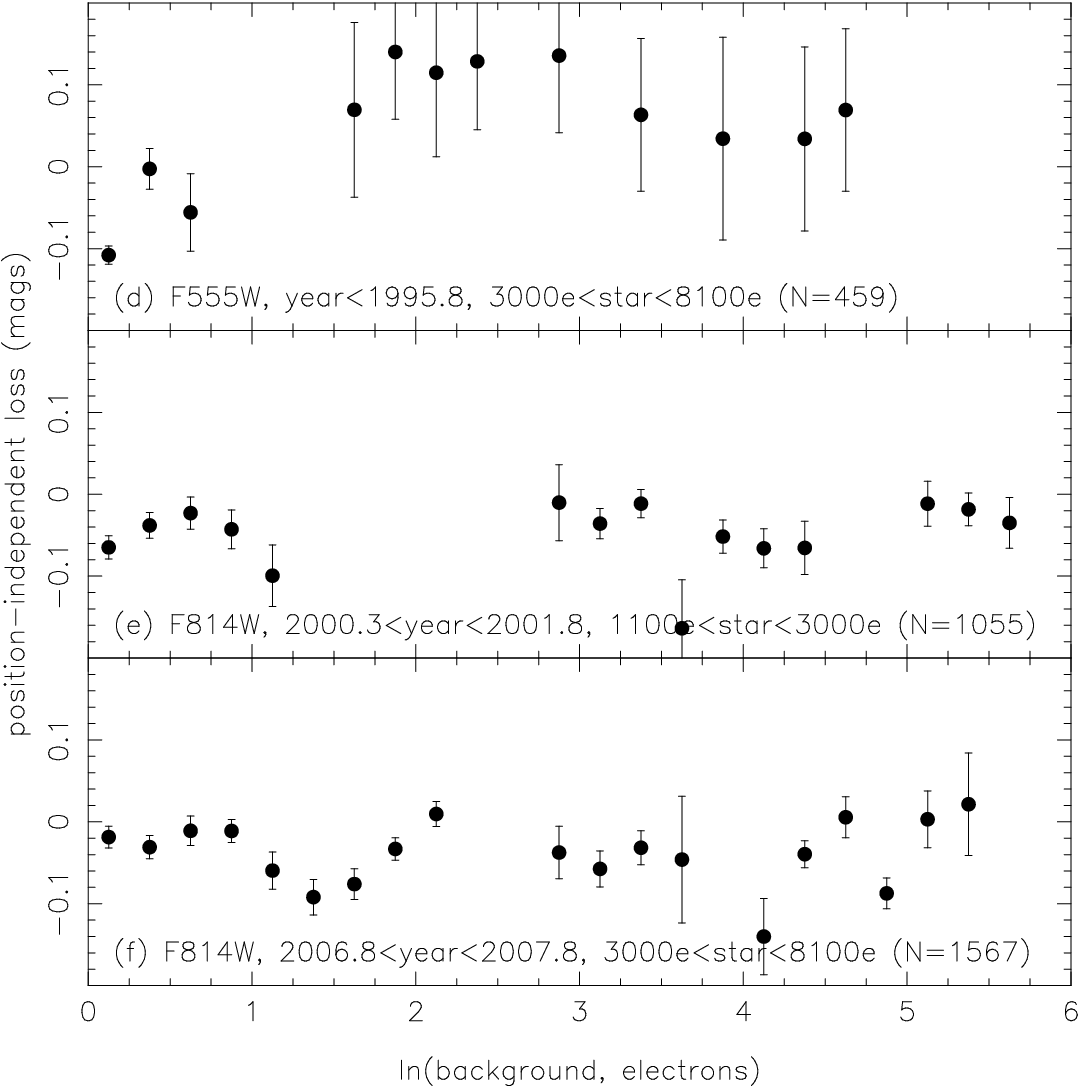}
\caption{Position-independent photometry errors, shown vs. star brightness (left) and background level (right). The data selections are shown in each panel. Note that, aside from offsets (zero point errors), no clear trends are visible.\label{fig_nopos}}
\end{figure}

In addition to CTE loss, there have also been many concerns regarding
photometric errors that are not position-dependent,
such as zero point offsets and the long-vs-short
anomaly \citep{cas98}.  Variations in the intercept of the fit
from Equation \ref{eq_linfit} vs. star brightness and vs. background level
are shown in Figure \ref{fig_nopos}.  (Note that position-independent
errors vs. observation date are merely time variations in the WFPC2
sensitivity, and are tracked extensively by the instrument team and thus
not addressed in this study.)

From the figure, one sees that no significant trends in the data are present.
Quantitatively, the data permit one to place a $1 \sigma$ upper limit of
0.025 magnitudes per dex of either star brightness or background level over
the ranges of each shown in the figure.

It should be noted that this finding not only dispels the presence of a
large long-vs-short anomaly, but also provides additional confidence
regarding the validity of the comparison between the ground-based standard
field and the WFPC2 observations.  Specifically, were any error as a
function of color present (which would masquerade as an error as a function
of magnitude, given the nature of this field), it would
have been seen in Figure \ref{fig_nopos}.  As no such error was observed,
the confidence in the validity of this comparison is enhanced.

What is measured, however, is a mean zero point offset relative to the
\citet{hol95b} calibrations.  Averaging the three plots of F555W data, the
mean residual (WFPC2 minus ground) is $\sim -0.06$ magnitudes, while
the mean for the three plots of F814W data is $\sim -0.03$ magnitudes.
Thus, a correction to the zero points will be needed in order to adequately
match the WFPC2 observations to the standard data.

\subsection{XCTE} \label{xcte}

\begin{figure}
\plottwo{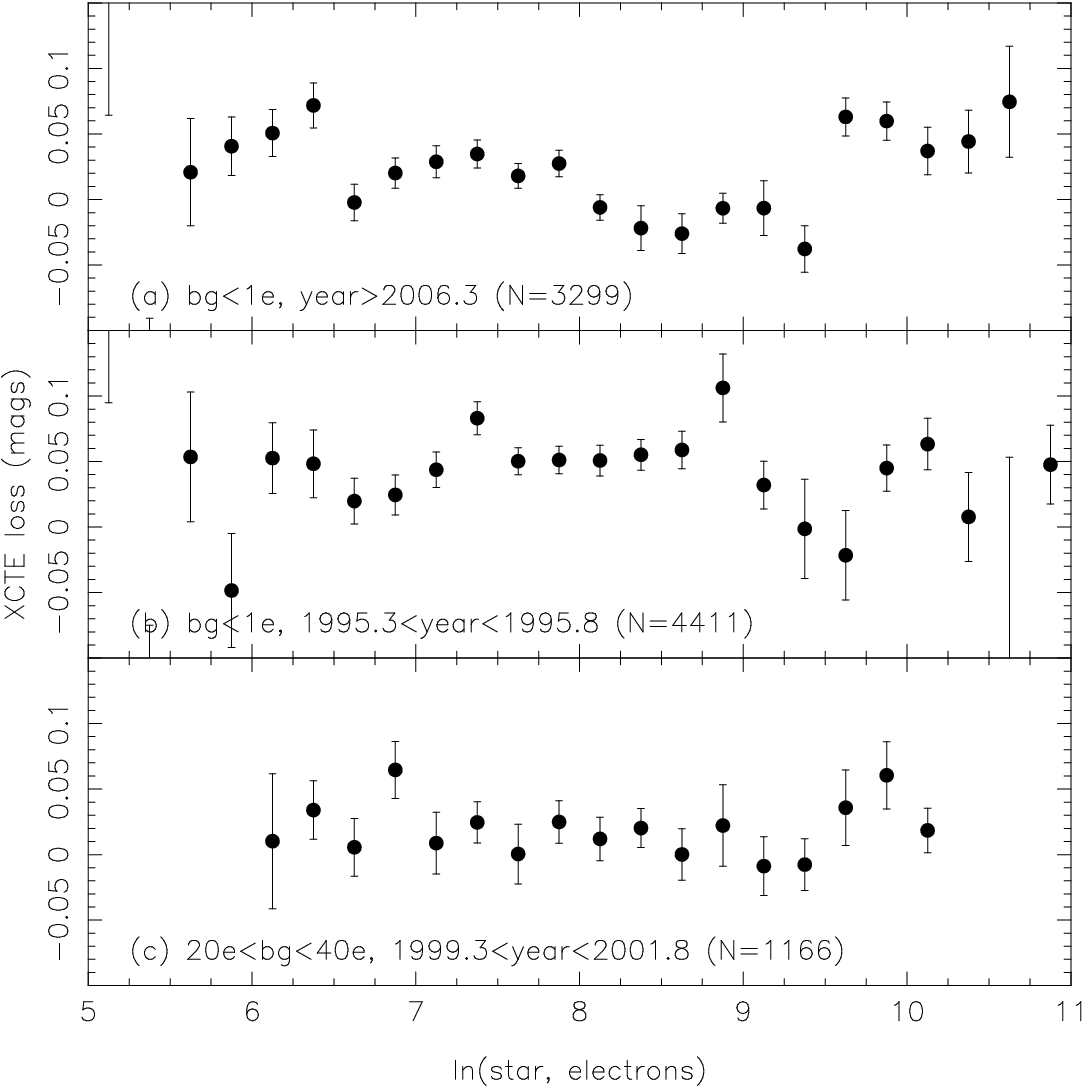}{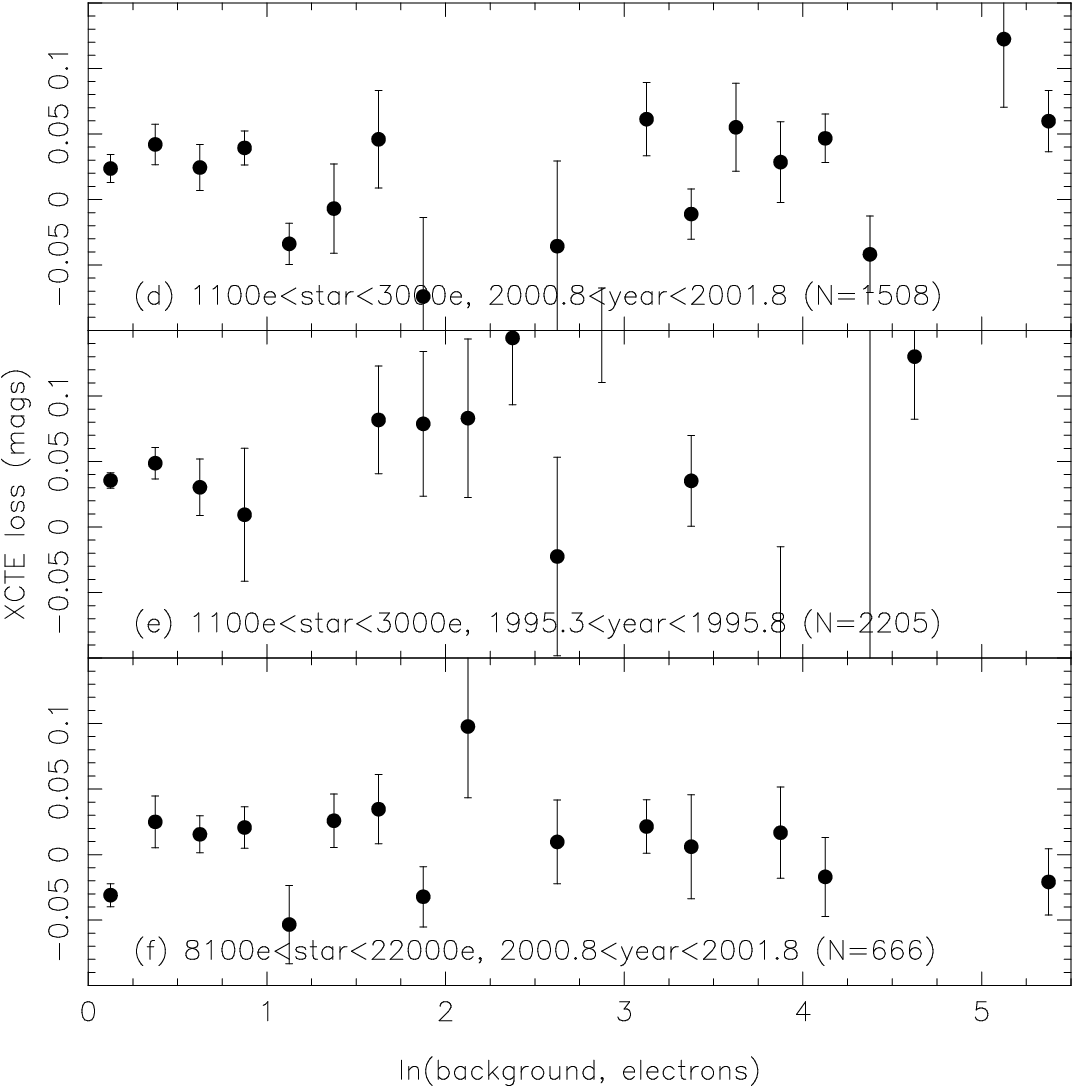}
\caption{XCTE losses,as a function of star brightness (left) and background level (right).  The data selections are shown in each panel. Due to the small size of XCTE losses, trends are not obvious from the figure.\label{fig_xcte}}
\end{figure}

While YCTE loss is easy to measure by eye and has grown significantly
over the years, XCTE loss has always had a small (though
measurable) effect on stellar photometry on WFPC2.  Figure \ref{fig_xcte}
shows XCTE loss as a function of star brightness and of background level.

As can be seen from the figure, dependencies of XCTE on the independent
variables is not obvious.  (This is also true of variations in XCTE as a
function of observation date, not shown in Figure \ref{fig_xcte}.)
Thus, determination of XCTE dependencies
will be left for statistical methods in Section 4.

\subsection{YCTE vs. Date} \label{cte_year}

\begin{figure}
\plotone{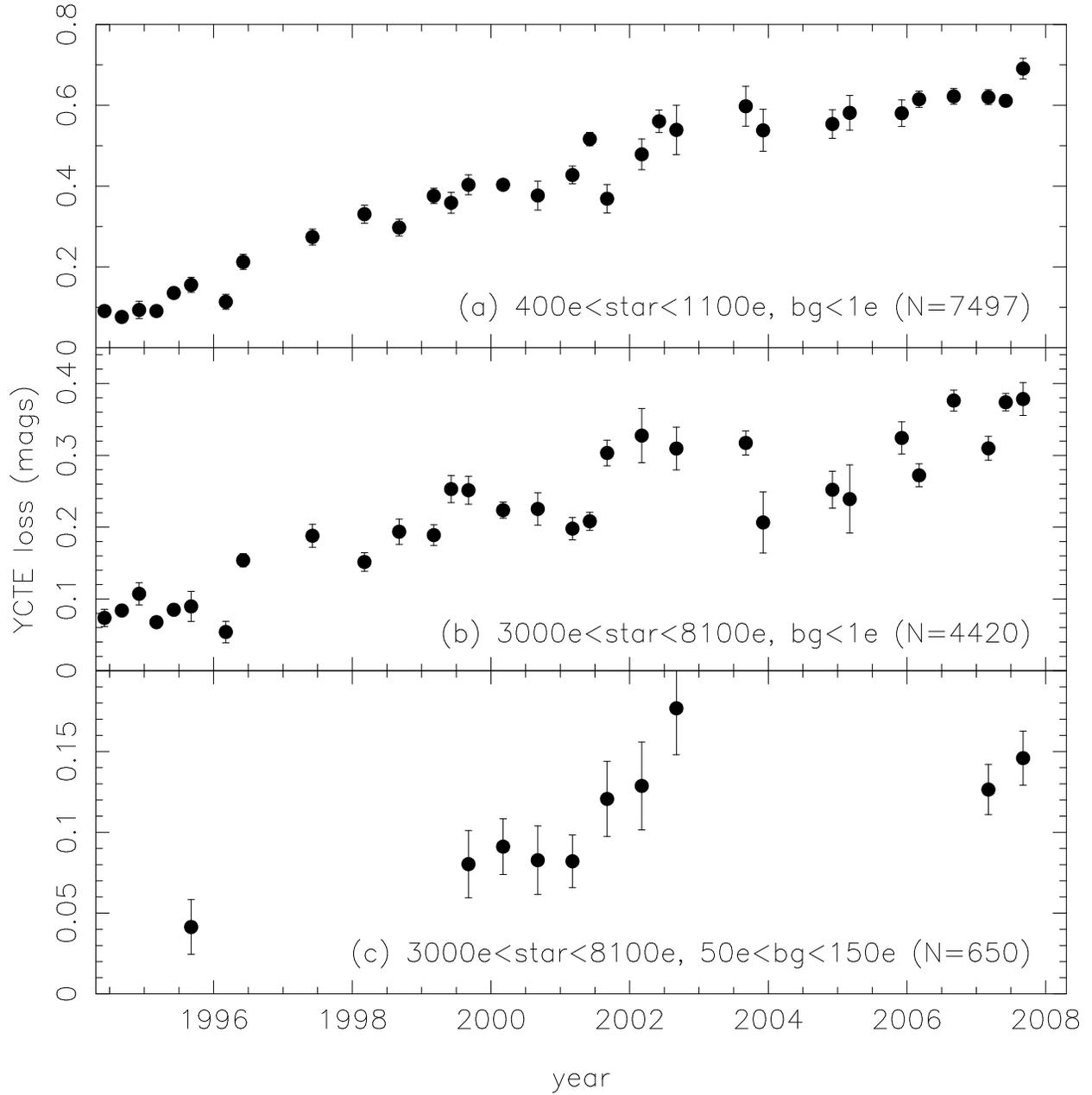}
\caption{YCTE losses, as a function of observation date.  Cuts in star brightness and background level are shown in the respective panels.\label{fig_ycte_yr0}}
\end{figure}

Figure \ref{fig_ycte_yr0} shows YCTE loss as a function of date of
observation.  To provide an indication of how this function is affected
by star brightness and background levels, three sets of data are shown:
faint stars on low background (panel a), bright stars on low
background (panel b), and bright stars on higher background (panel
c).  While these samples do not represent the extremes of the
data shown in Table \ref{tab_stars}, these are the largest baselines
containing sufficient numbers of stars to see clear trends.

An examination of panels (a) and (b) of Figure \ref{fig_ycte_yr0} indicates
that, despite nearly a factor of two difference in YCTE loss between faint
and bright stars in recent observations (both on low backgrounds), both
sets of data show similar YCTE losses during the earliest observations.
Specifically,
quadratic fits to the curves give YCTE losses of $0.054 \pm 0.005$ magnitudes for
faint stars (panel a) and $0.058 \pm 0.004$ magnitudes for bright stars (panel b)
for an observation date of 1994.3 (20 April 2004).
This indicates that the two curves are not merely scaled versions of each
other.
It should also be noted that the same
dependence on observation date is seen in both curves, such that panels (a)
and (b) both fall onto the curve:
\begin{displaymath}
\Delta m = 0.056 + (\Delta \hbox{yr}-0.027 \Delta \hbox{yr}^2) \times f(\hbox{CTS}),
\end{displaymath}
where $\Delta \hbox{yr}$ is the year minus 1994.3.

The data for higher background levels are not nearly as complete, due to
few high-background observations of this field made prior to mid-1998.  The data in
Figure \ref{fig_ycte_yr0}c are consistent with both the same intercept at
1994.3, as well as a lower one.  In other words, determination of whether
the functional form should be
\begin{equation}
\Delta mag = 0.056 + (\Delta \hbox{yr}-0.027 \Delta \hbox{yr}^2) \times f(\hbox{CTS}) \times f(\hbox{BG}) \label{eq_bginside}
\end{equation}
or
\begin{equation}
\Delta mag = [ 0.056 + (\Delta \hbox{yr}-0.027 \Delta \hbox{yr}^2) \times f(\hbox{CTS}) ] \times f(\hbox{BG}) \label{eq_bgoutside}
\end{equation}
must wait until the analysis of the full data set is made.  It should be
noted that the calibration data used by \citet{hol95a} showed a strong
background dependence of YCTE loss for observations obtained in 1994: 0.04
magnitudes of loss (at $Y=800$) for measurements with background levels
$\la 20-30$ electrons, 0.02 magnitudes for background levels $\sim 30 -
\sim 250$ electrons, and zero for higher background levels.  Thus,
the functional form in Equation \ref{eq_bgoutside} is to be preferred unless
the present data set demonstrate otherwise.

\subsection{YCTE vs. Star Brightness} \label{cte_cts}

\begin{figure}
\plotone{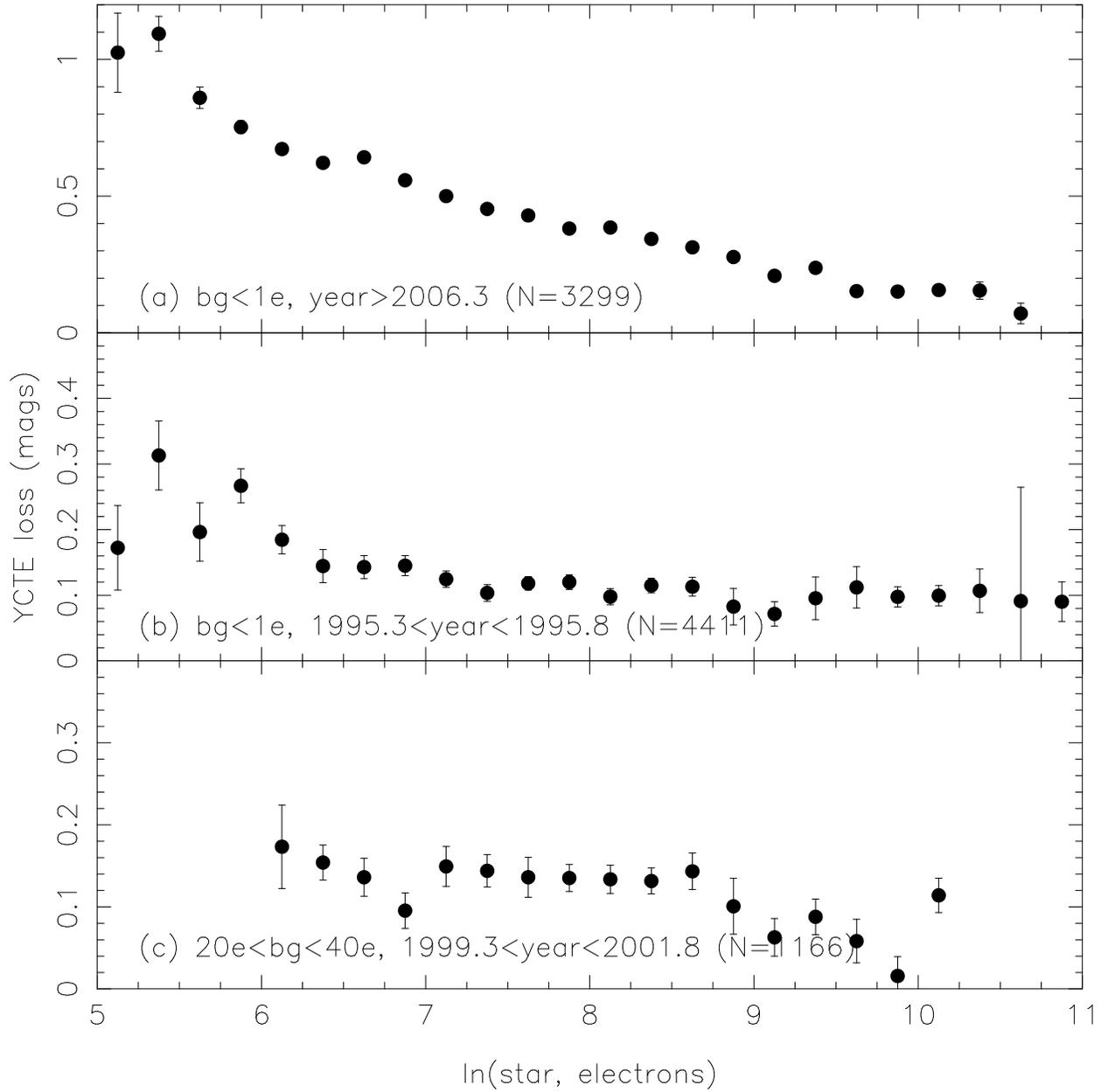}
\caption{YCTE losses, as a function of star brightness.  Cuts in background level and observation date are shown in the respective panels.\label{fig_ycte_cts0}}
\end{figure}

Figure \ref{fig_ycte_cts0} shows YCTE loss as a function of star brightness,
with data selected to show recent observations with low background (panel
a), old observations with low background (panel b), and
intermediate-age observations with higher background (panel c).

The trend from panels (a) and (b) is that the slope of the curve decreases
as star brightness increases, which suggests the use of an exponential for
the CTE variation as a function of brightness.

Considering only points for which YCTE loss is less than 0.5 magnitudes
(see Section \ref{new1} for a discussion of the treatment of large
CTE losses) and using the results of the previous section, both curves can
be fitted with the formula:
\begin{displaymath}
\Delta m = 0.056 + 1.2 (\Delta \hbox{yr}-0.027 \Delta \hbox{yr}^2) \times e^{-0.43 \ln(\hbox{\hbox{CTS}})}.
\end{displaymath}

Effects of background on the YCTE loss vs. counts will be discussed in
Section \ref{cte_bg}.

\subsection{YCTE vs. Background} \label{cte_bg}

\begin{figure}
\plotone{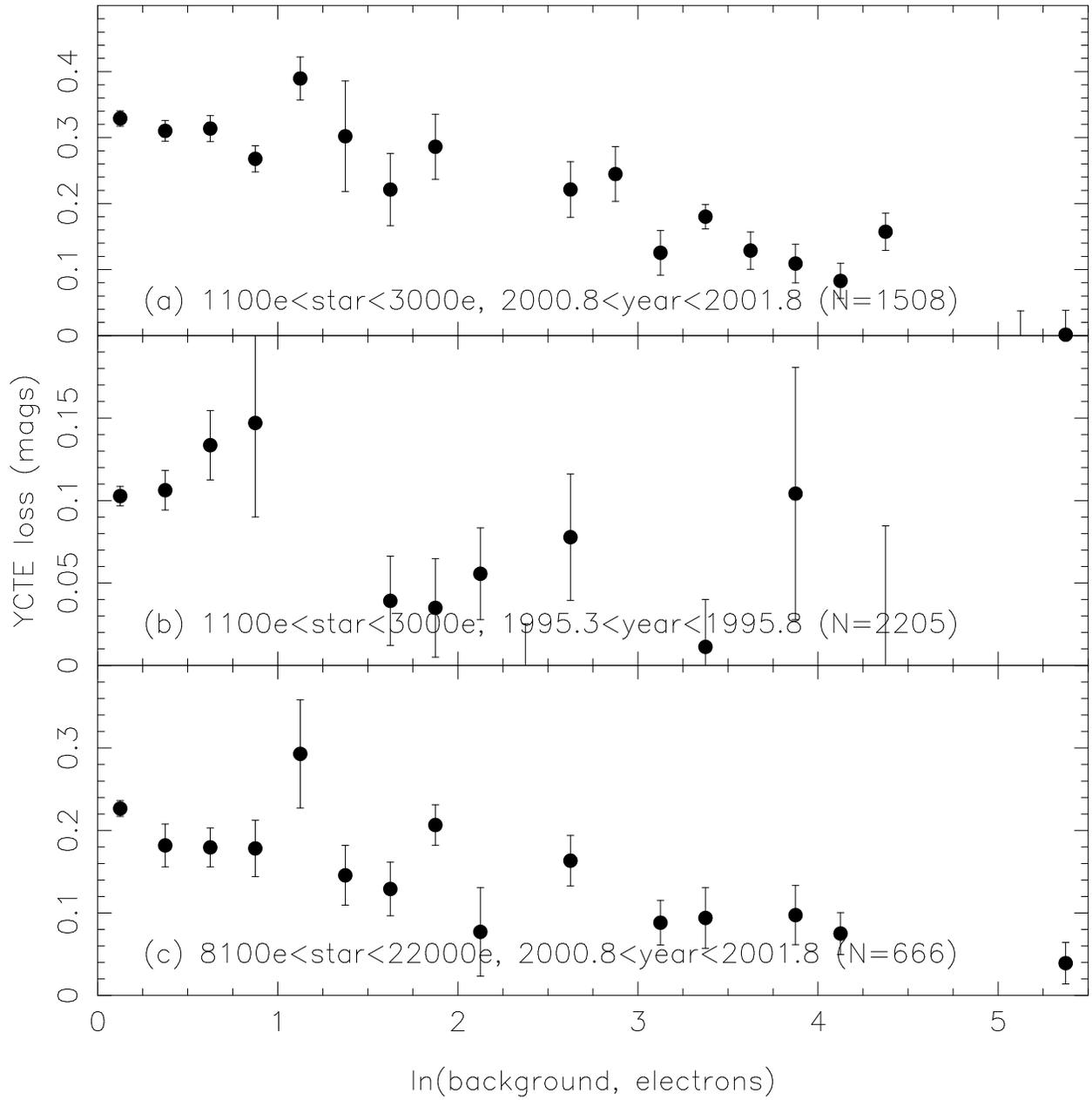}
\caption{YCTE losses, as a function of background level.  Cuts in star brightness and observation date are shown in the respective panels.\label{fig_ycte_bg0}}
\end{figure}

The final independent variable considered in this analysis is the
background level.  Figure \ref{fig_ycte_bg0} shows CTE loss as a function
of background, for faint stars observed in roughly 2001 (panel a),
faint stars observed in 1995 (panel b), and bright stars observed
in 2001 (panel c).  As with D00 and D02, a ``softened''
background was used, which equals $\sqrt{1+\hbox{BG}^2}$ electrons
(and is 1 if the
measured background level is negative).  This softened background compresses
the lowest background levels, making it easier to find a functional form for
fitting CTE loss vs. background.

The YCTE plots in Figure \ref{fig_ycte_bg0} do not
show the curvature characteristic of an exponential decay, suggesting that
a line be used instead to fit the YCTE dependence on background.
Another significant feature of the background dependence, although
the data used to generate Figure \ref{fig_ycte_bg0} contain insufficient stars to make
this evident, is that the slope of the relation is shallower for bright
stars than it is for faint stars.
Thus, the function used to fit the background dependence is
\begin{displaymath}
f(\hbox{BG}) = \max{[( 1 - c_1 \ln{\hbox{BG}} + c_2 ( \ln{\hbox{BG}} - c_3 ) \ln{\hbox{CTS}} ), c_4]}.
\end{displaymath}

\subsection{Charge Loss During Readout} \label{new1}

When CTE loss removes a significant fraction of a star's overall brightness,
one must take into account the fact that the CTE loss rate itself changes
as the star is read out.  Using the customary terminology of losses as
fractions of star brightness lost (as opposed to number of
electrons lost) per pixel read out,
the CTE loss rate will increase during readout since the
loss rate increases as the star becomes fainter.

The exact relation can be determined fairly simply.
Let the charge lost per pixel transfered be expressed as
a differential equation with the form
\begin{equation} \label{eq_diffeq}
\frac{dx}{dy} = - a - b e^{-cx},
\end{equation}
where $x$ is the natural logarithm of the star's brightness; and
$a$, $b$, and $c$ are constants (or functions of background level and
observation date, which are both constant during the readout process).
If the star's brightness were to remain essentially unchanged during
readout (i.e., CTE loss is a small fraction of the star's brightness),
the loss per pixel traversed would be constant as well.  This would result
in a magnitude loss $\Delta m$ of
\begin{equation} \label{eq_simple}
\Delta m = 1.086 Y (a+be^{-cx}),
\end{equation}
where $Y$ is the star's Y position on the chip.  The multiplier of $1.086$
converts
from differences in natural logarithm to differences in magnitude.  Note that
this general form is seen in the literature correction equations
(Whitmore 1998; Stetson 1998; Whitmore et al. 1999; D00).

However, for very faint stars, the amount of CTE loss can be significant,
and thus one cannot safely assume that CTE loss remains
constant during readout.  Solving Equation \ref{eq_diffeq}, the magnitude
loss is given by
\begin{equation} \label{eq_complex}
\Delta m = \frac{1.086}{c} \ln [ e^{acy}(1+\frac{b}{a}e^{-cx}) - \frac{b}{a}e^{-cx} ],
\end{equation}
where $x$ is the natural logarithm of the measured counts (after any loss
due to readout).  Note that, if $acy \ll 1$, this equation simplifies
to Equation \ref{eq_simple}.
For larger values of $acy$,
this produces a smaller CTE correction than Equation \ref{eq_simple}.

It should be noted that a CTE correction derived from data including
extremely faint stars
should implicitly include this effect, as the lower CTE loss would
have been in the measurements and thus in the correction.  However, most CTE
corrections (such as that presented in this paper) are based only on brighter stars,
due to the practical difficulty in obtaining sufficiently accurate
photometry of extremely faint stars.  Thus, the
brightness dependence in the CTE
corrections is measured using brighter stars for which Equations
\ref{eq_simple} and \ref{eq_complex} produce the same result.  Assuming
that the brightness dependence is correct (given the extrapolation involved,
it may not be), the form of the correction from Equation \ref{eq_complex}
will be required to correctly model CTE loss of faint stars.

\begin{figure}
\plottwo{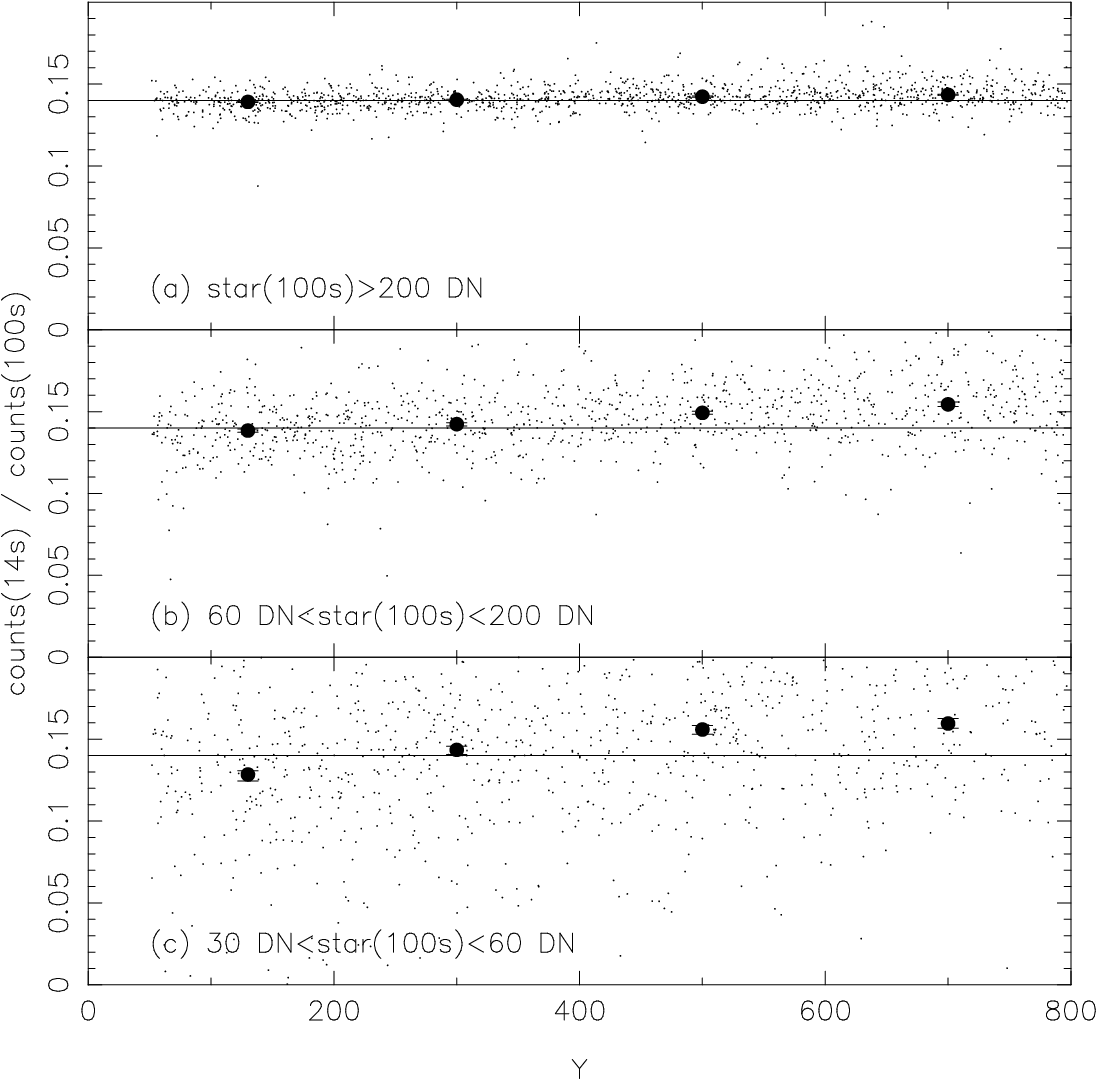}{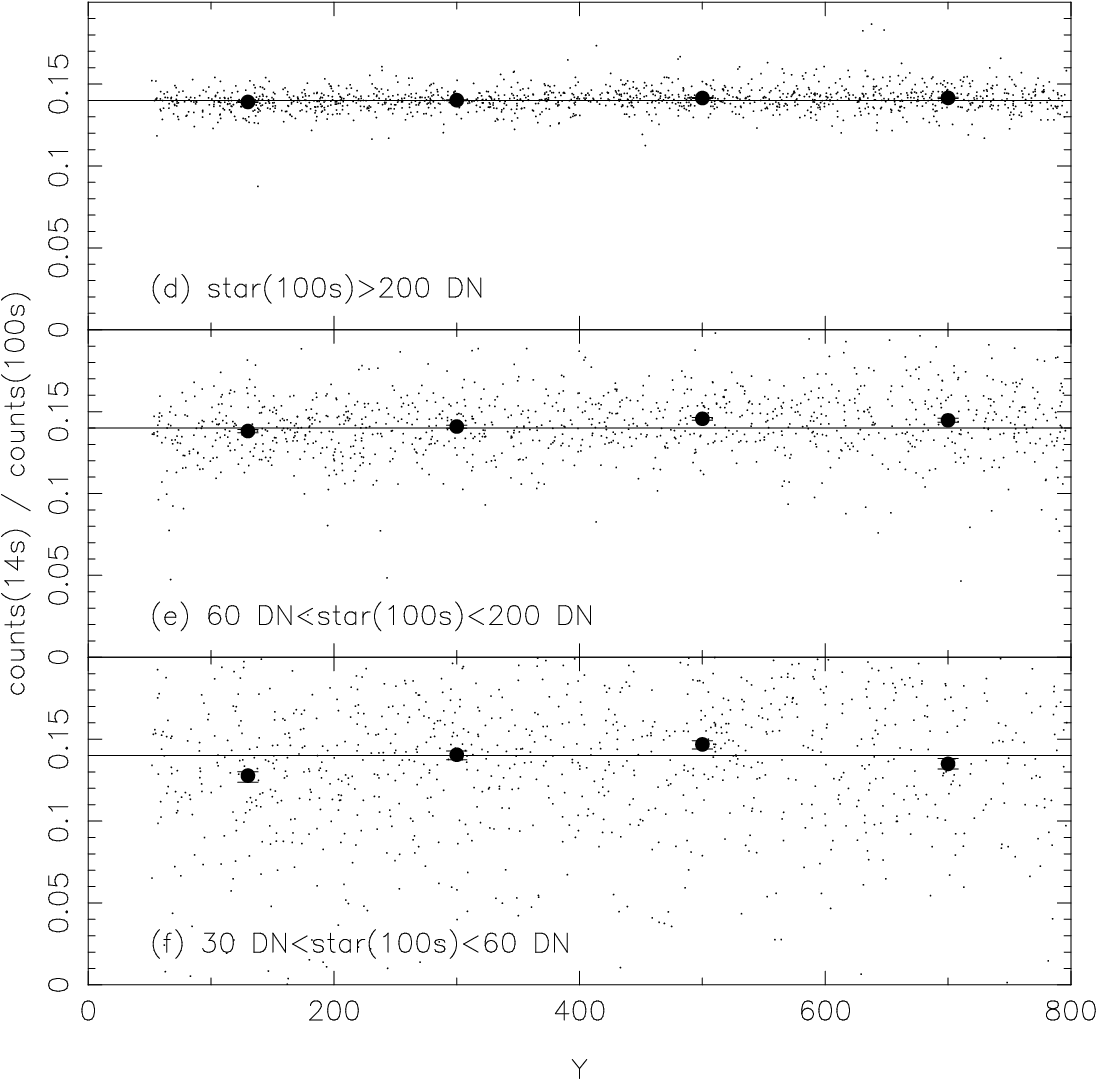}
\caption{Ratio of counts from 14-second exposures to 100-second exposures, after applying CTE correction. The form of the CTE correction used in panels a-c fails to account for the change in CTE loss during readout, while that used in panels d-f account for this effect.  The horizontal line is at 0.14, the expected ratio.  The brightest stars (panels a and d) correspond to brightness levels of the calibration data used in this study (400 electrons and brighter in the short exposures).  Note that CTE loss is overcorrected for the faintest stars when not accounting for CTE loss rate changes during readout.\label{fig_veryfaint}}
\end{figure}

This effect is illustrated in Figure \ref{fig_veryfaint}.  To produce this
figure, observations of the $\omega$ Cen standard field, taken on 17 August
2000 as part of calibration program GO-8447, were analyzed.
The observations consisted of one 100-second F814W
exposure and two 14-second F814W exposures (as well as other exposures
using different filters and preflashes) at each of two pointings.  For
each pointing, HSTphot was used to obtain photometry for the 100-second
exposure as well as the average of the 14-second exposures.  Figure
\ref{fig_veryfaint} shows ratios of the 14-second to 100-second
count levels after applying CTE corrections of the forms in Equations
\ref{eq_simple} and \ref{eq_complex}.
The horizontal line in the figure is at a ratio of 0.14, which is what
would be expected in the absence of CTE loss.

As is clear from the figure, the form of the CTE correction that accounts
for CTE loss during readout significantly reduces systematic errors in the
correction.  This also bolsters confidence that the CTE corrections
presented in the next section are valid for stars significantly fainter
than those used for to compute the corrections themselves.

\section{CTE Corrections}

Guided by the analysis from the previous section, a solution was made for
zero points and CTE corrections.  Several fits were made,
with variations on the form of the CTE correction.  The form that produced
the best fit to the data was
\begin{equation}
\frac{dm}{dx} = c_1 \times \hbox{\hbox{BG}}^{-c_2} \times ( 1 + c_3 \hbox{DATE} )
\end{equation}
and
\begin{eqnarray}
\frac{dm}{dy} & = & \{ c_4 + c_5 [(\hbox{DATE}-c_6) - c_7 (\hbox{DATE}-c_6)^2] \hbox{CTS}^{-c_8} \} \nonumber \\
& & \times \max{ [( 1 - c_9 \ln{\hbox{BG}} + c_{10} \ln{\hbox{BG}} \ln{\hbox{CTS}} + c_{11} \ln{\hbox{CTS}} ) , c_{12}] },
\end{eqnarray}
where BG is the background level, CTS is the star's brightness, DATE is
the observation date, and the left-hand sides are the magnitude loss per
pixel read out in the $x$ and $y$ directions.  Note that this form is not
exactly that from Equation \ref{eq_diffeq}, as the star's brightness appears
twice.  However, as long as $c_{10}$ is sufficiently small that changes to
the star's brightness during readout do not significantly affect the BG term,
the solution in Equation \ref{eq_complex} can be used.

After obtaining the best fit to the CTE correction parameters (and zero point
differences), the ``standard'' photometry was redetermined from the corrected
WFPC2 data.  The rationale is that 1184 WFPC2 data sets ought to produce
photometry that is superior to that of the original calibration data.  In
addition, while the results from section 3.1 indicate that issues with the
ground-based photometry (such as errors in color corrections) do not
generate any significant error in the WFPC2 vs. ground comparison, this will
eliminate any chance that even a small error in the ground-based data would
affect the measured CTE corrections.

\begin{figure}
\plottwo{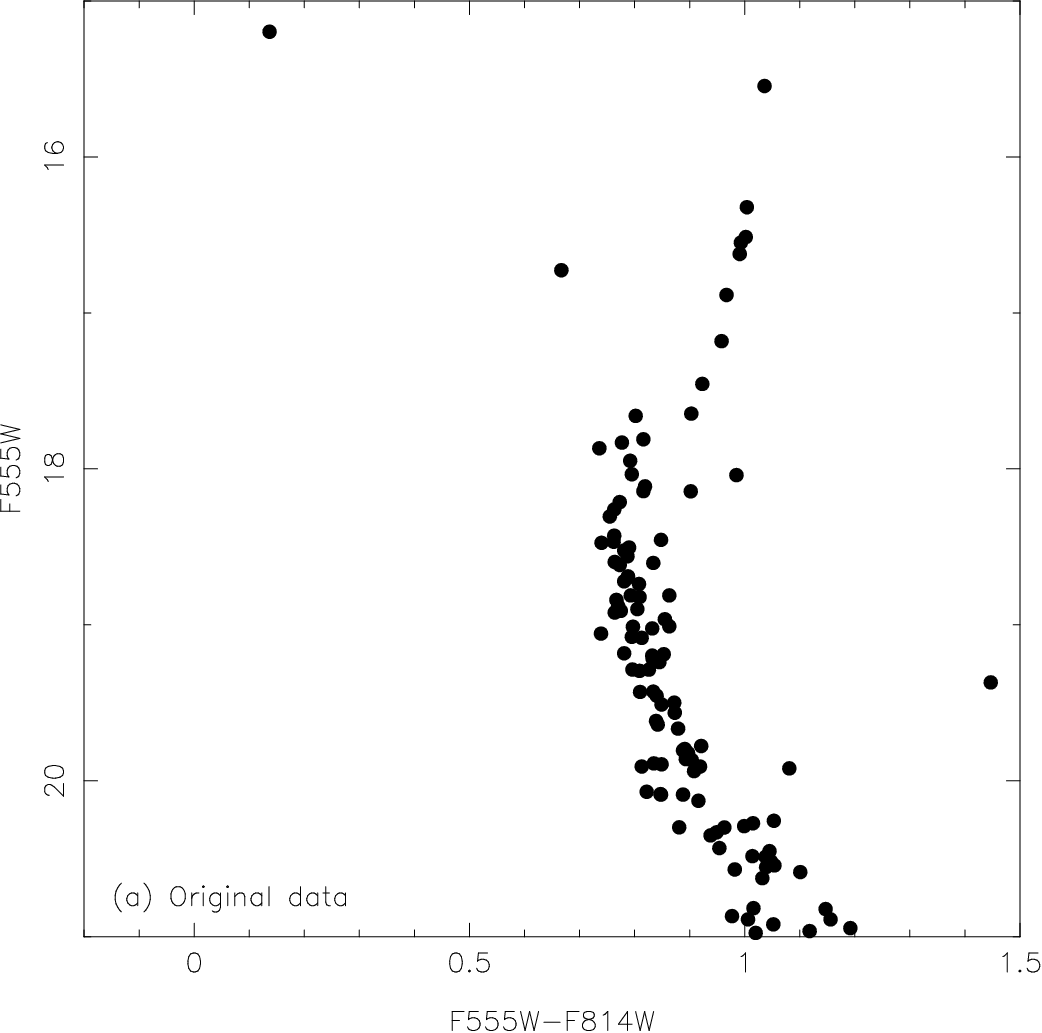}{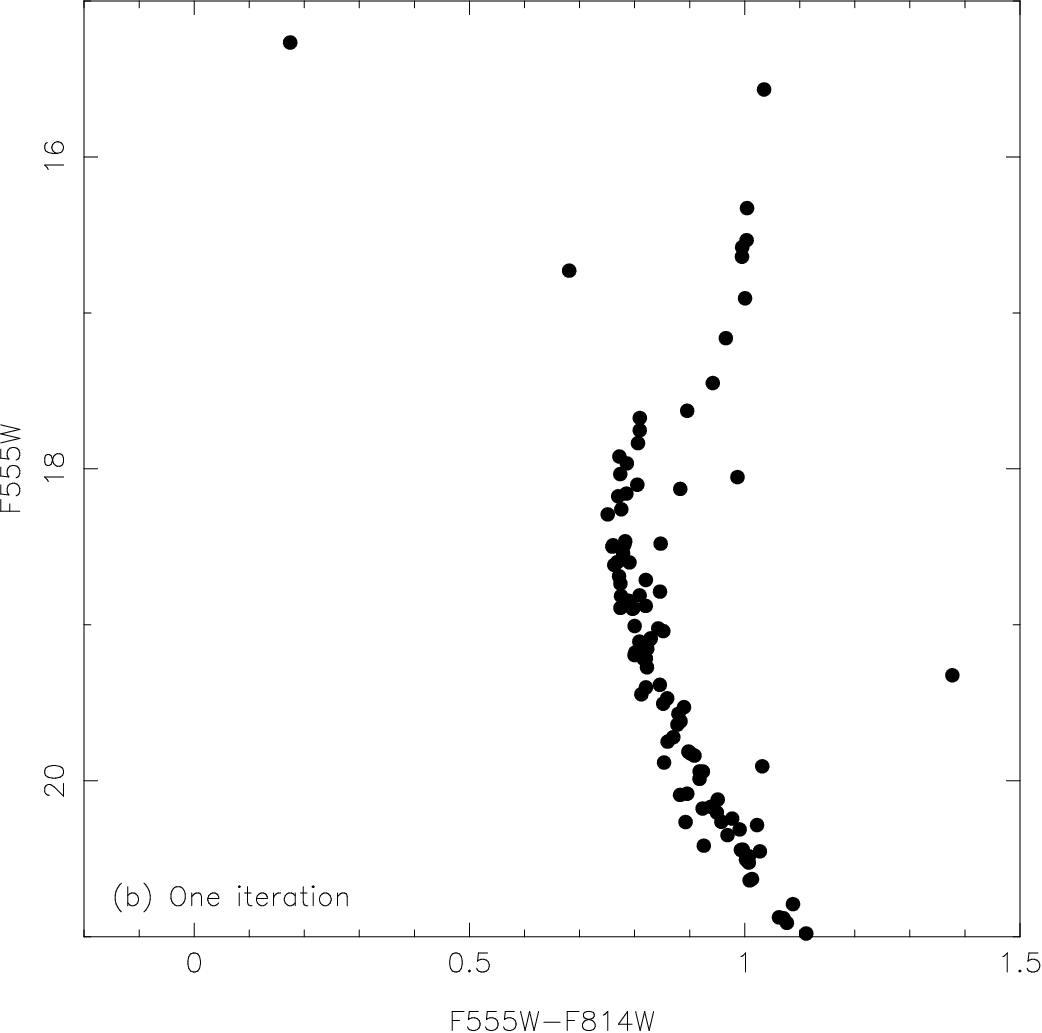}
\plottwo{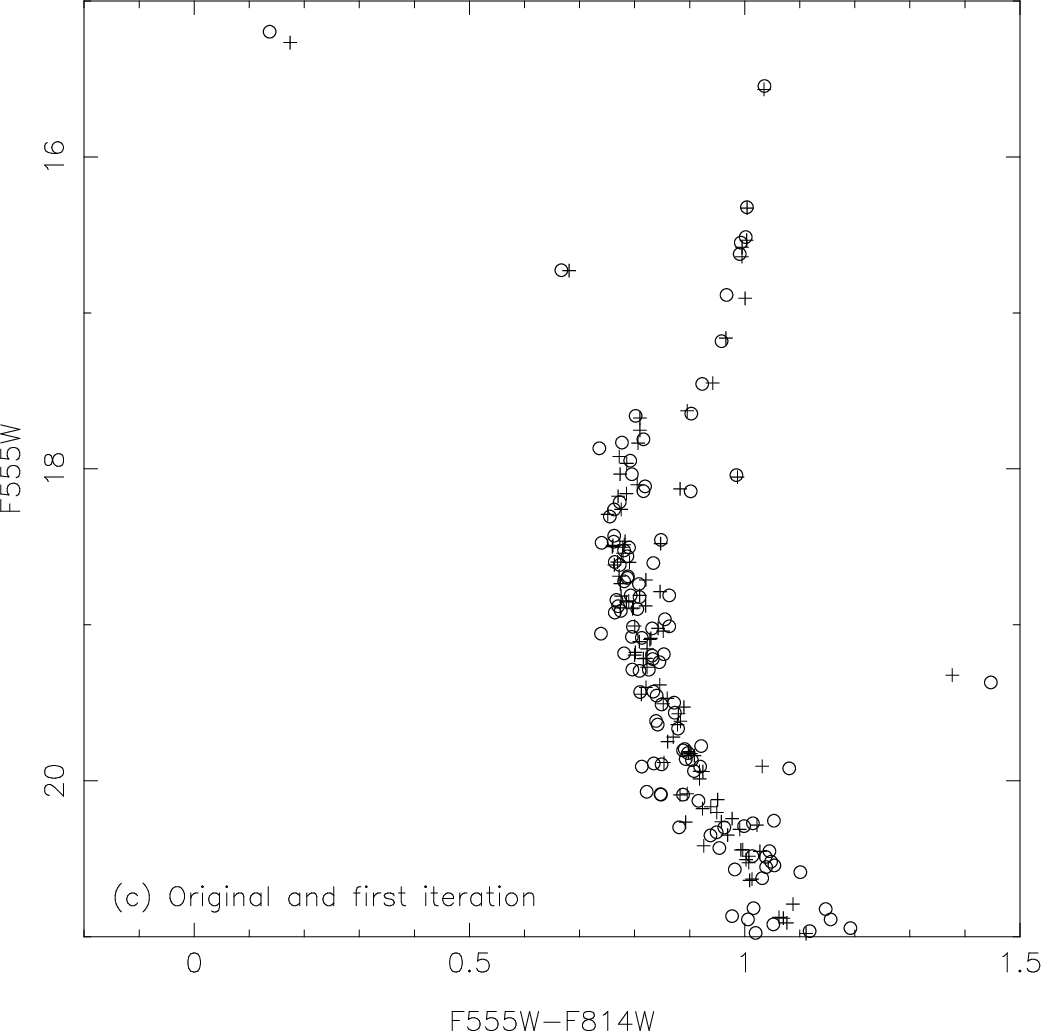}{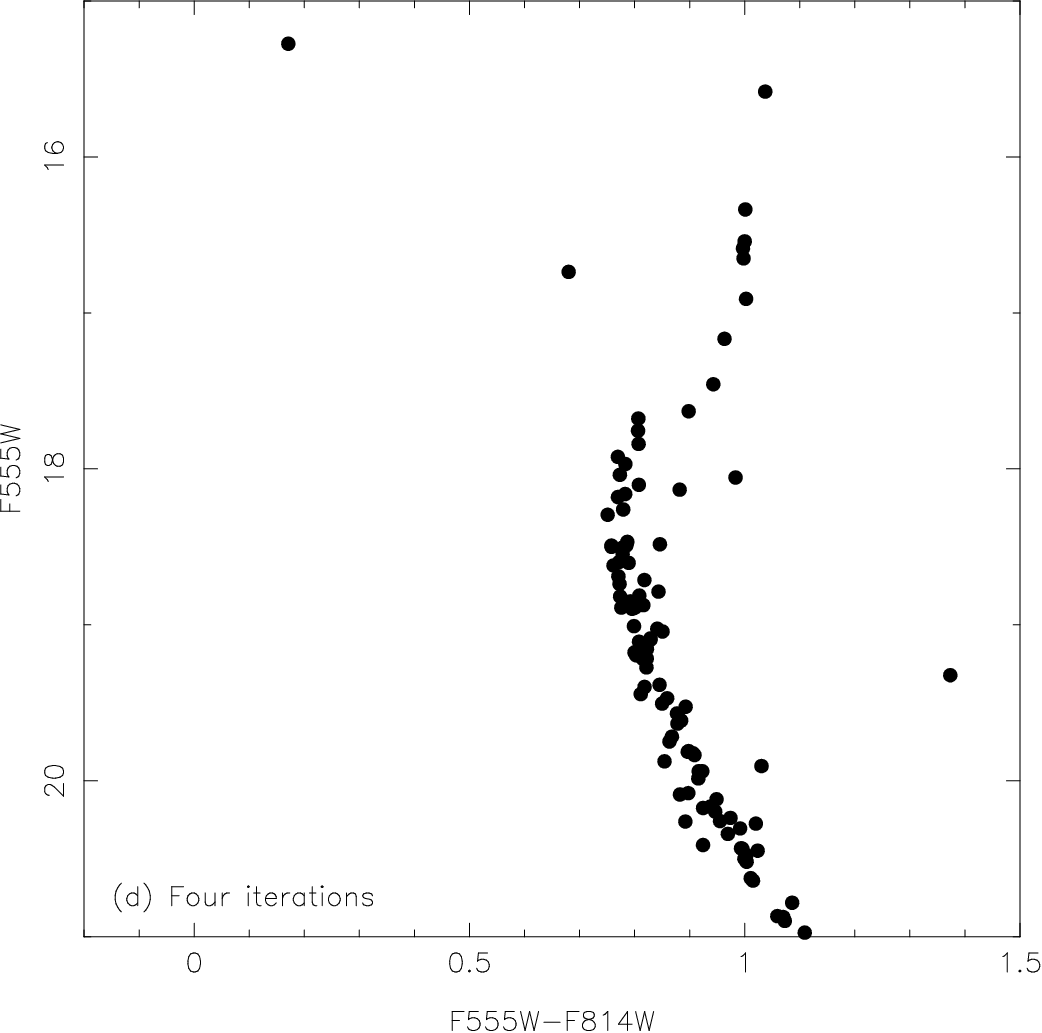}
\caption{Standard photometry used in determining CTE corrections.  Panel (a) shows the original photometry of \citet{wal94}, transformed into F555W and F814W instrumental magnitudes.  Panel (b) shows the values obtained from CTE-corrected WFPC2 photometry, and panel (c) shows the first two panels overplotted.  Note that the sequences lie overlaid, indicating that no significant photometry biases are introduced by using the WFPC2 data for a subsequent CTE solution.  Finally, panel (d) shows the values from CTE-corrected WFPC2 photometry, after four iterations of re-computing standard magnitudes.  The largest magnitude difference between this and panel (b) is 0.01.\label{fig_cmds}}
\end{figure}

The only negative aspect to this use of WFPC2-measured standard data to
measure WFPC2 CTE losses is that the form of the CTE solution will implicitly
become part of the standard photometry used to derive the CTE solution --
a sort of circular logic.  However, Figure \ref{fig_cmds} shows the original
and updated CMDs, and it is clear that a huge improvement in the standard
photometry was obtained by making this improvement without introducing
biases.  Were one to repeat this
process -- measure the CTE losses from the new standard photometry, determine
yet newer standard photometry, redetermine the CTE losses, and so on --
the standard photometry would change by at most 1\%, as shown in Panel (d) of
Figure \ref{fig_cmds}.  Thus, the CTE solution recommended here is the one
with a single iteration on the standard magnitudes.  Naturally, the
suggested zero points are based on the original ground-based standards.

To verify that the recommended solution is truly the best fit,
alternative solutions made the following modifications to the functional form:
\begin{itemize}
\item adding a brightness dependence to the XCTE correction.  This produced a null result (i.e., the dependency was zero to within the uncertainties).
\item using an exponential function of background (instead of a power law) in the XCTE correction.  This produced a small degradation in the quality of the fit.  Due to the small size of the XCTE correction, this is of minimal significance.
\item adopting a background dependence of the form in Equation \ref{eq_bginside} (instead of that in Equation \ref{eq_bgoutside}).  Due to the lack of high-background observations of this field early in WFPC2's lifetime, there was no statistical difference between the quality of the fits.  It was therefore decided to adopt the form from Equation \ref{eq_bgoutside}, whose solution for early observations is in strong agreement with the CTE characterization of \citet{hol95a}.
\item moving the count dependence also outside the brackets in Equation \ref{eq_bgoutside}.  This was is the most common form in literature CTE corrections, including D00 and D02.  This produced a significantly worse fit, and can be ruled out.
\item using a power law for the YCTE background dependence.  This was also used in most literature corrections, including D00 and D02.  This also produced a significantly worse fit, and can be ruled out.
\end{itemize}

The recipe for correcting CTE loss is as follows.  Given a star brightness CTS, background level BG (both in electrons), observation date DATE (in MJD), and star position on the image $X$ and $Y$, the following sequence of calculations will provide the $XCTE$ and $YCTE$ losses, both in magnitudes.
\begin{equation}
\mbox{lbg} = \frac{1}{2} \ln(\mbox{BG}^2 + 1 ) - 1
\end{equation}
\begin{equation}
\mbox{yr} = \frac{\mbox{DATE} - 49461.9}{365.25}
\end{equation}
\begin{equation}
XCTE = 0.0077 e^{-0.50 \mbox{lbg}} (1+0.10\mbox{yr}) \frac{X}{800}
\end{equation}
\begin{equation}
\mbox{lct} = \ln(\mbox{CTS}) + 0.921 XCTE - 7
\end{equation}
\begin{equation}
c = 0.958 (\mbox{yr}-0.0255 \mbox{yr}^2) e^{-0.450 \mbox{lct}}
\end{equation}
\begin{equation}
YCTE = 2.41 \ln[(1+c)e^{0.02239 \max{(1.0-0.201 \mbox{lbg}+0.039 \mbox{lbg} \mbox{lct}+0.002 \mbox{lct},0.15)} (Y/800)}-c]
\end{equation}
Note that the offsets of 1 for lbg and 7 for lct were put in place for
numerical stability, and do not affect the solution itself.

The zero points of the four primary filters (F439W, F555W, F675W, and
F814W) were measured independently for each chip and gain setting (thus, a
total of eight zero points per filter).  For other filters, the relatively
small number of available data sets required that their zero points be
characterized with a
single offset relative to the zero point of the nearest primary filter.

Zero points for both gain settings are provided in Tables \ref{tab_zp7}
and \ref{tab_zp14}.  Instrumental magnitudes can be computed using
\begin{equation}
m_{WFPC2} = -2.5 \log(\hbox{counts/sec}) + ZP - XCTE - YCTE,
\end{equation}
where the counts/second rate is measured within an aperture whose diameter is
one arcsecond.

The $BVRI$ transformations of \citet{hol95b} have also been
modified to account for the new zero points, and are provided in Table
\ref{tab_xform}.  (Note that the color terms have been adopted without
modifications, and are reprinted here for convenience.)
Standard $BVRI$ magnitudes can be computed with the following equation:
\begin{equation}
SMAG = -2.5 \log(\hbox{counts/sec}) + ZP + T1 \times SCOL + T2 \times SCOL^2 - XCTE - YCTE.
\end{equation}
For instances in which a single set of parameters does not describe the
relation sufficiently
well for all colors, the Cmin and Cmax columns of Table \ref{tab_xform} show the color range for
which the table row is valid.
Again, the count rate is measured within an aperture whose diameter is one
arcsecond.

\subsection{Verification}

\begin{figure}
\plottwo{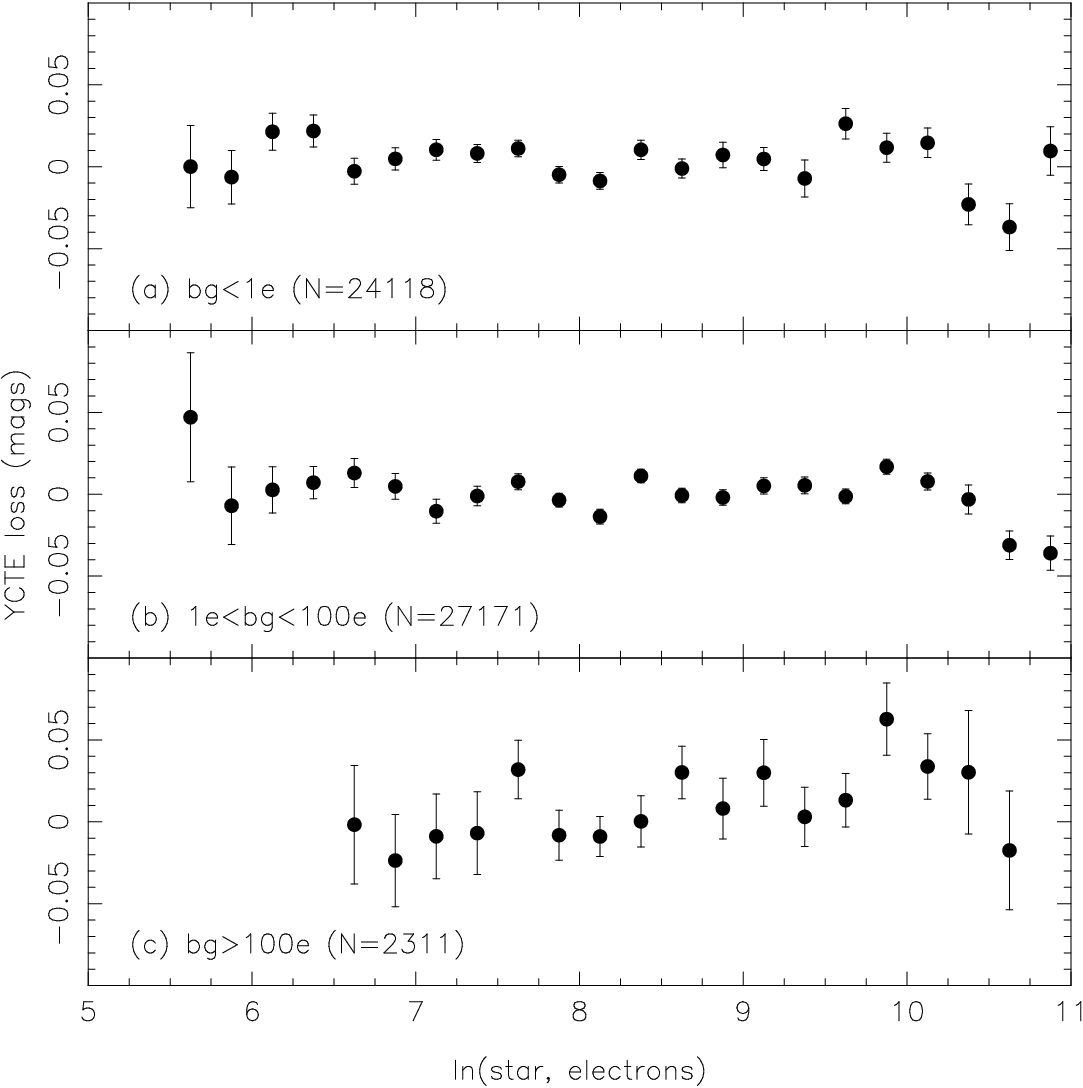}{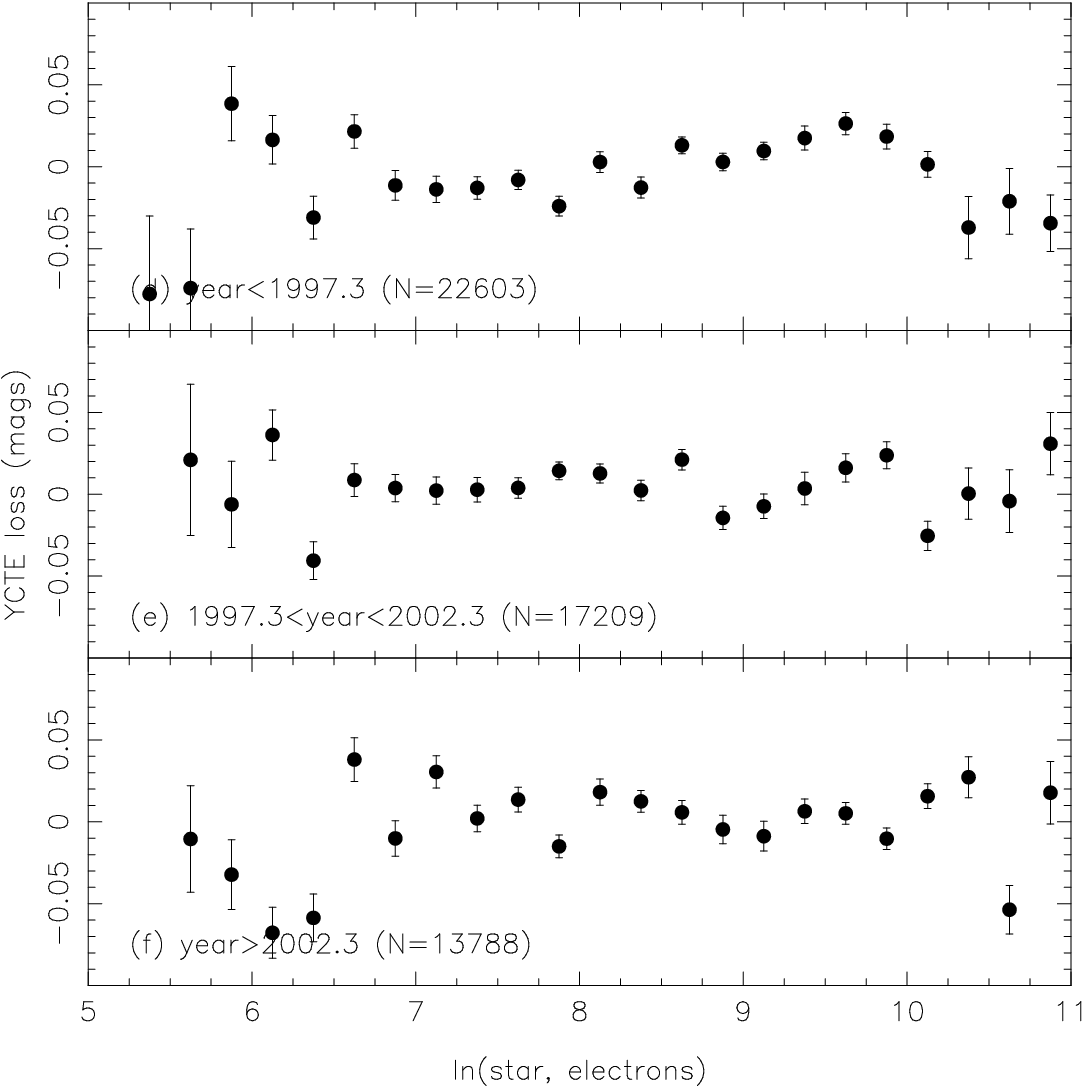}
\caption{Residual YCTE losses (measured minus fit), as a function of star brightness.  The left panels (a-c) show three different data samples selected by background level; the data in the right panels (d-f) are selected by observation date.\label{fig_cts_last}}
\end{figure}

\begin{figure}
\plottwo{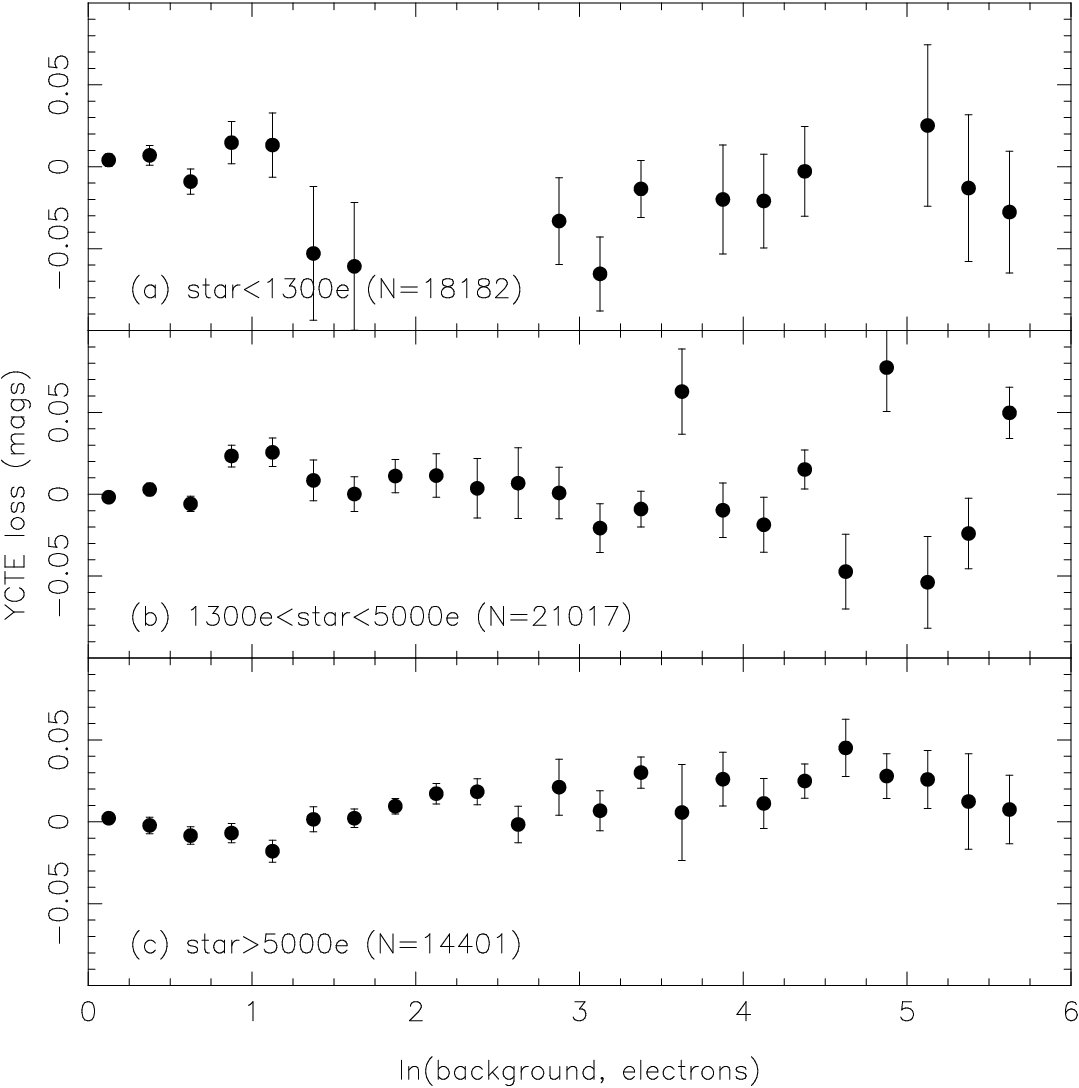}{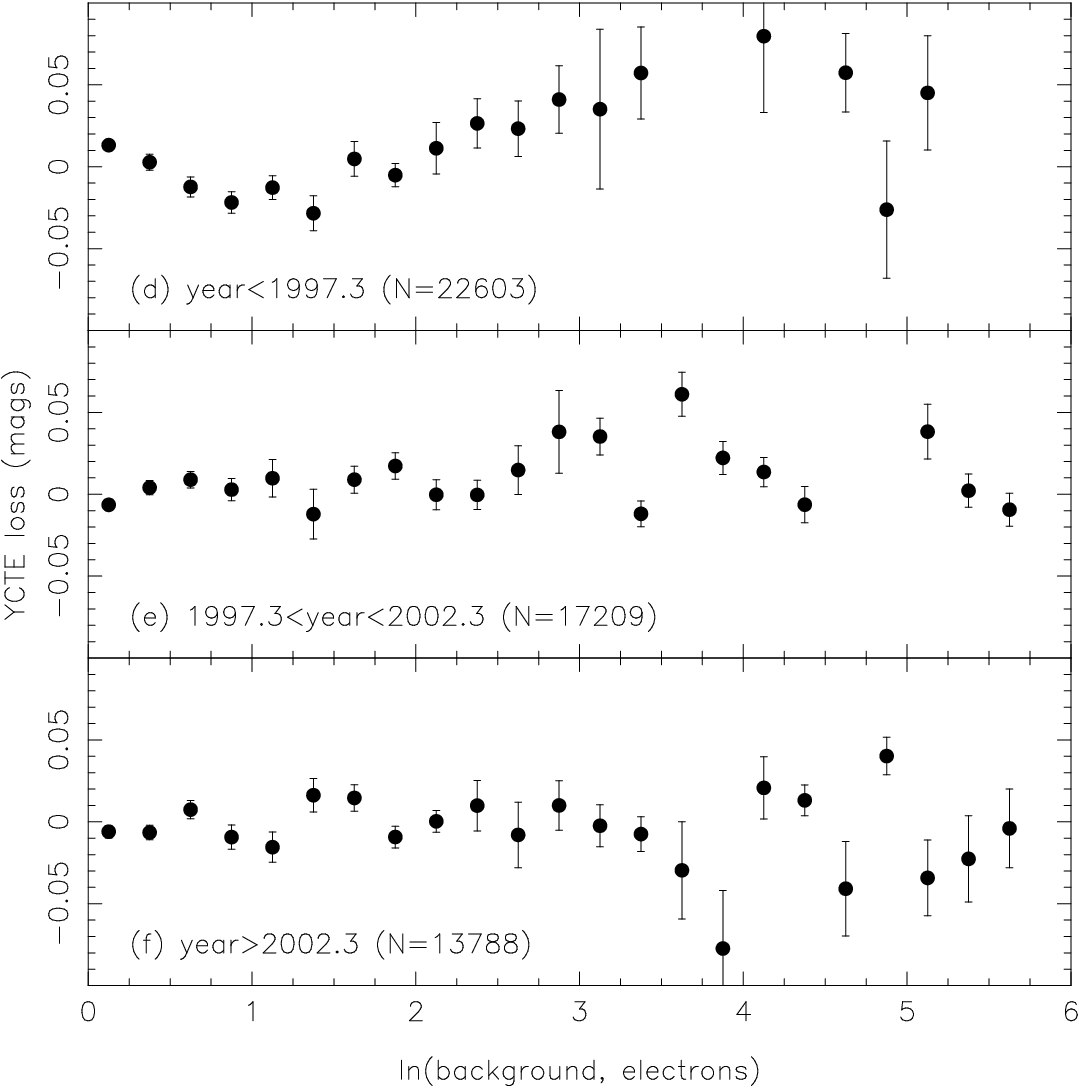}
\caption{Residual YCTE losses (measured minus fit), as a function of background level.  The left panels (a-c) show three different data samples selected by star brightness; the data in the right panels (d-f) are selected by observation date.\label{fig_bg_last}}
\end{figure}

\begin{figure}
\plottwo{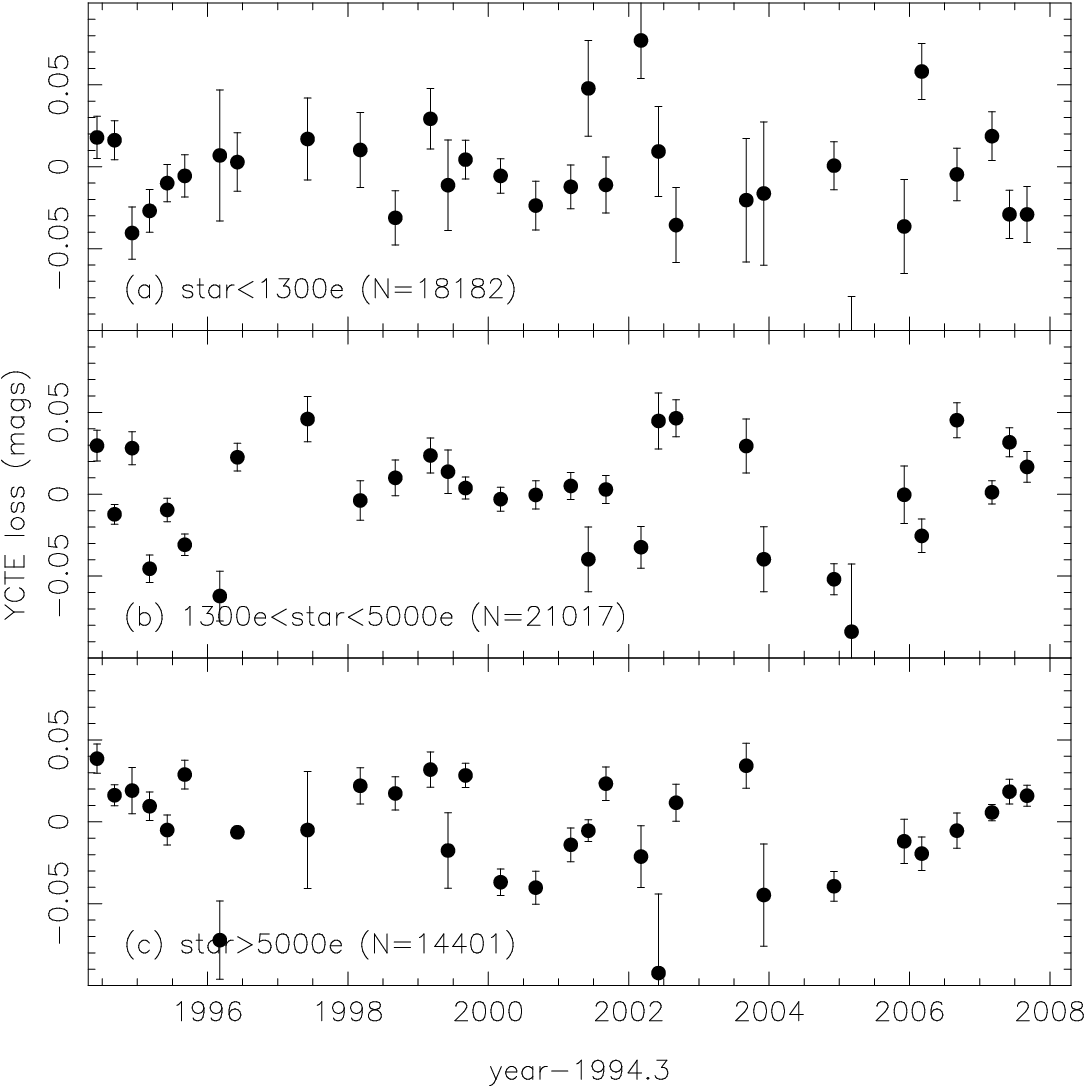}{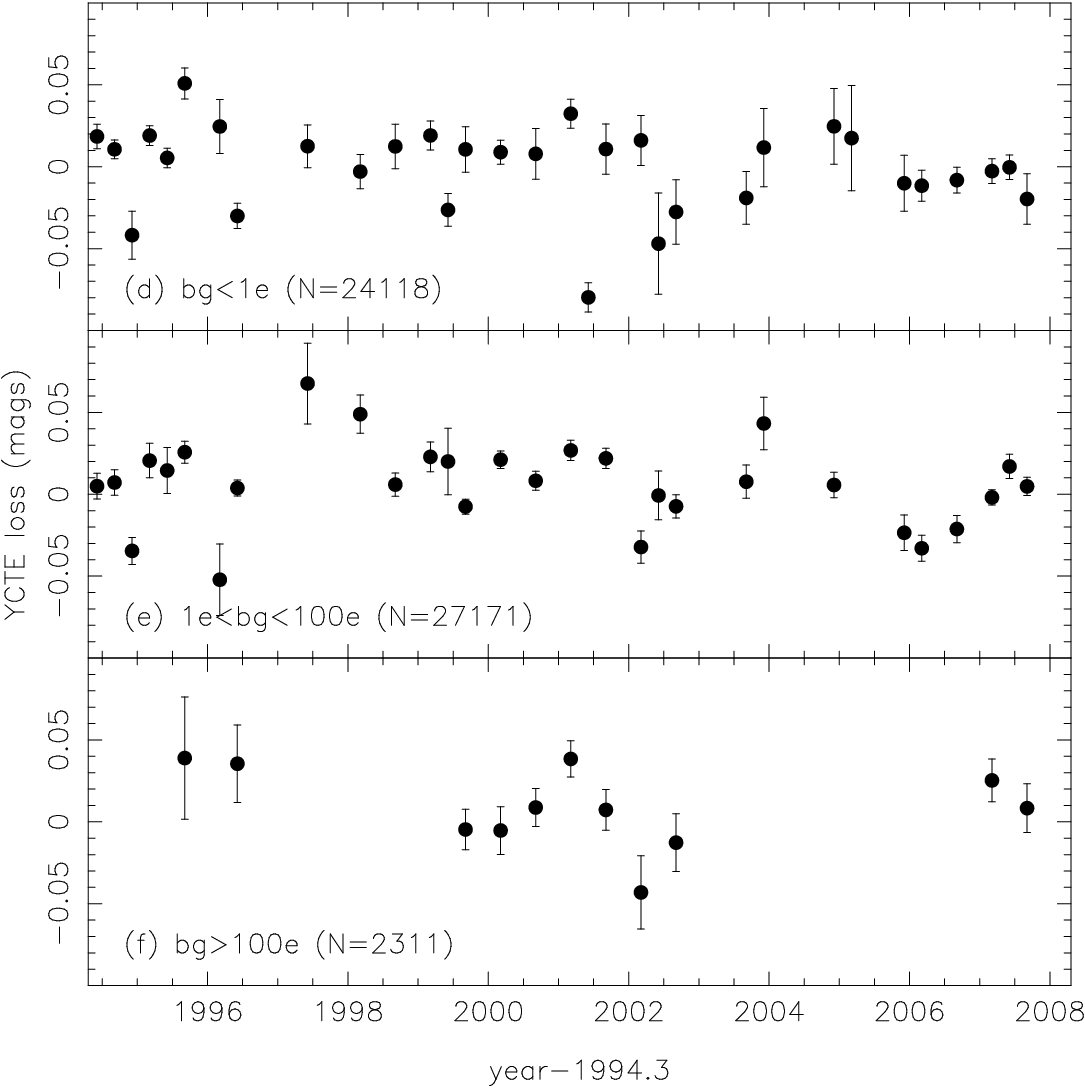}
\caption{Residual YCTE losses (measured minus fit), as a function of observation date.  The left panels (a-c) show three different data samples selected by star brightness; the data in the right panels (d-f) are selected by background level.\label{fig_yr_last}}
\end{figure}

To verify the accuracy to which the data are being corrected, Figures
\ref{fig_cts_last}, \ref{fig_bg_last}, and \ref{fig_yr_last} show the
YCTE residuals after zero point and CTE corrections have been applied.
These plots are generated similarly to
Figures \ref{fig_ycte_yr0}$-$\ref{fig_ycte_bg0}, although larger data
selections have been used.

With the possible exceptions of the background dependence for early
observations (Figure \ref{fig_bg_last}d), no significant trends remain in
the data.  It should also be noted that the upward trend in
Figure \ref{fig_bg_last}d is based on very little data; only 544 star
measurements (of the 22603
used for the data in this panel) had background levels of 20 or more
electrons.

\section{Consequences of the New Formulae}

Comparing the CTE corrections from this work with those of D02,
the CTE corrections agree to within 0.02 magnitudes for over 80\% of the
star measurements.
Stars for which the current corrections are more than 0.02 magnitudes larger
fall primarily into two categories: bright stars on a bright background,
and dim stars (brightnesses of a few hundred electrons) on
backgrounds of a few electrons.
Not surprisingly, stars for which the current corrections are more than
0.02 magnitudes smaller also fall into two categories: bright stars
on a low background (one electron or less), and dim stars on bright
backgrounds.

\begin{figure}
\plottwo{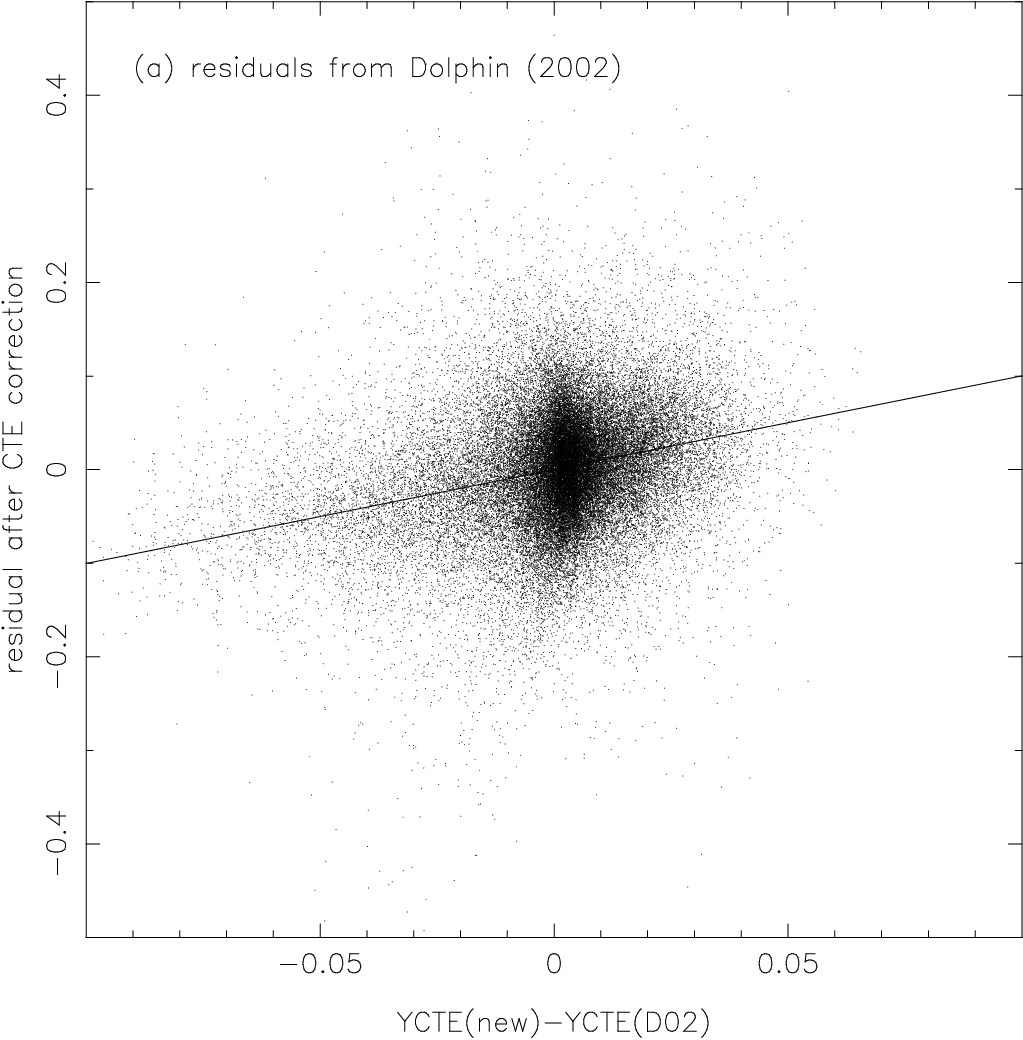}{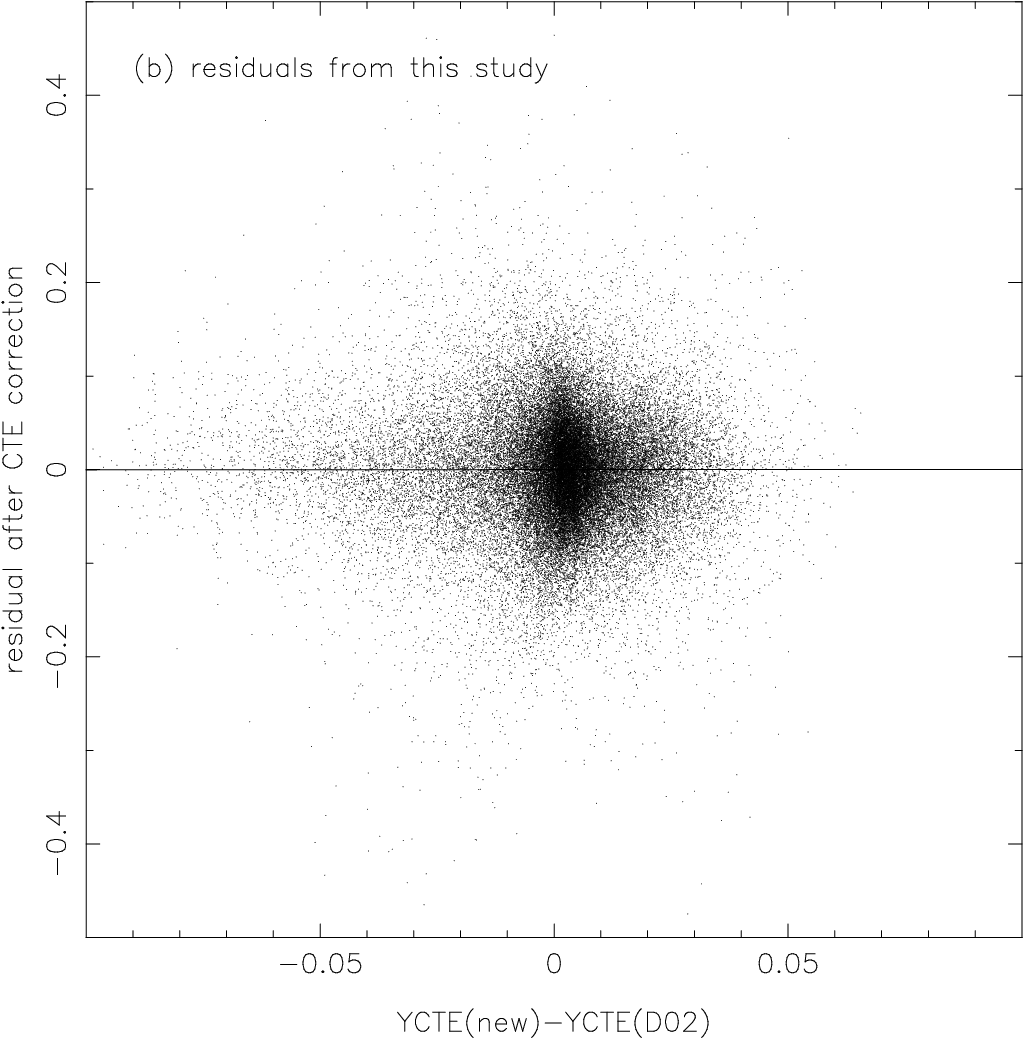}
\caption{A comparison of the D02 CTE corrections with those from this study.  Residuals after applying the D02 CTE correction (left panel) and the CTE correction from this study (right panel) are plotted against difference between the two corrections.  The lines in each panel show the best fit to the data.\label{fig_ctecomp}}
\end{figure}

The reason for the difference is the term allowing the background
correction slope to change as a function of star brightness, which was not
present in the D02 CTE equation.  To verify that the new corrections indeed
fit the data better, Figure \ref{fig_ctecomp} shows the residuals for
both sets of CTE corrections, plotted versus the difference between the
CTE corrections.  The relation between CTE difference and residual is
clear in Figure \ref{fig_ctecomp}a, and in fact has a slope of one.
Using the corrections from this paper, no trend of residual as a function of
CTE difference exists (Figure \ref{fig_ctecomp}b).

Given that typical science exposures have background levels higher than
is typical in calibration exposures, this result can have
a significant on the analysis of WFPC2 data.

\section{Summary}

The study presented here revisits the topic of WFPC2 charge transfer
inefficiency, building on results of previous work (D00, D02).
This examination is overdue, as the existing CTE characterizations are
several years old and thus application to current WFPC2 observations
requires significant extrapolation of the calibration data to the current
epoch.

In addition to incorporation of more recent calibration data, several other
enhancements were made over the previous work.  Most significantly, it was seen
that the effects of star brightness and background level cannot be treated
independently.  Making this assumption gave reasonable CTE corrections for
stars whose brightness or background level was typical of calibration data.
However, for higher background levels, the older CTE corrections will undercorrect bright
stars and overcorrect faint stars.  For bright stars, the resulting systematic
error can significantly exceed the random error from photon noise.

Other significant changes from the earlier work include accounting for the
change in CTE loss during readout, an improved functional form for the
background dependence, and a count dependence that begins at zero at the
beginning of the WFPC2 mission.

As additional calibration data become available, updated WFPC2 calibration
data will be made available
on the author's website\footnote{http://purcell.as.arizona.edu/wfpc2\_calib/}.

\acknowledgments

Based on observations made with the NASA/ESA Hubble Space Telescope, obtained from the data archive at the Space Telescope Science Institute.  Support for this work was provided by NASA through grant number AR-11244 from the Space Telescope Science Institute, which is operated by AURA, Inc., under NASA contract NAS 5-26555.

{\it Facilities:} \facility{HST (WFPC2)}.

\clearpage

\clearpage

\clearpage

\clearpage

\clearpage

\clearpage

\clearpage

\clearpage

\clearpage

\clearpage

\clearpage

\clearpage

\begin{deluxetable}{lccc}
\tabletypesize{\scriptsize}
\tablecaption{Simple Statistics of Stellar Measurements Used for Calibration\label{tab_stars}}
\tablewidth{0pt}
\tablehead{\colhead{Value} & \colhead{Brightness\tablenotemark{a}} & \colhead{Background\tablenotemark{a}} & \colhead{Epoch}}
\startdata
minimum & 106   & 0    & 29 April 1994 \\
16\%    & 666   & 0.16 & 25 December 1994 \\
median  & 2226  & 1.16 & 15 June 1999 \\
84\%    & 8658  & 9.2  & 24 November 2005 \\
maximum & 59971 & 2229 & 21 August 2007 \\
\enddata
\tablenotetext{a}{Values given in electrons}
\end{deluxetable}

\clearpage

\begin{deluxetable}{lcccc}
\tabletypesize{\scriptsize}
\tablecaption{Zero Points for GAIN=7\label{tab_zp7}}
\tablewidth{0pt}
\tablehead{\colhead{Filter} & \colhead{PC} & \colhead{WFC2} & \colhead{WFC3} & \colhead{WFC4}}
\startdata
F380W  & $21.000 \pm 0.009$ & $21.012 \pm 0.005$ & $21.012 \pm 0.005$ & $20.996 \pm 0.005$ \\
F410M  & $19.360 \pm 0.009$ & $19.373 \pm 0.006$ & $19.373 \pm 0.005$ & $19.356 \pm 0.005$ \\
F439W  & $20.874 \pm 0.008$ & $20.886 \pm 0.004$ & $20.886 \pm 0.004$ & $20.870 \pm 0.004$ \\
F450W  & $21.970 \pm 0.009$ & $21.982 \pm 0.005$ & $21.982 \pm 0.005$ & $21.966 \pm 0.005$ \\
F467M  & $19.895 \pm 0.009$ & $19.908 \pm 0.006$ & $19.908 \pm 0.005$ & $19.891 \pm 0.005$ \\
F547M  & $21.618 \pm 0.003$ & $21.627 \pm 0.002$ & $21.630 \pm 0.003$ & $21.609 \pm 0.002$ \\
F555W  & $22.508 \pm 0.003$ & $22.517 \pm 0.001$ & $22.520 \pm 0.002$ & $22.499 \pm 0.001$ \\
F569W  & $22.181 \pm 0.003$ & $22.190 \pm 0.002$ & $22.193 \pm 0.003$ & $22.172 \pm 0.002$ \\
F606W  & $22.850 \pm 0.003$ & $22.859 \pm 0.002$ & $22.862 \pm 0.003$ & $22.841 \pm 0.002$ \\
F622W  & $22.329 \pm 0.009$ & $22.340 \pm 0.006$ & $22.338 \pm 0.007$ & $22.327 \pm 0.006$ \\
F675W  & $22.033 \pm 0.007$ & $22.044 \pm 0.004$ & $22.042 \pm 0.004$ & $22.030 \pm 0.003$ \\
F702W  & $22.432 \pm 0.008$ & $22.443 \pm 0.004$ & $22.442 \pm 0.005$ & $22.430 \pm 0.004$ \\
F785LP & $20.635 \pm 0.003$ & $20.668 \pm 0.003$ & $20.670 \pm 0.003$ & $20.642 \pm 0.003$ \\
F791W  & $21.441 \pm 0.004$ & $21.473 \pm 0.004$ & $21.475 \pm 0.004$ & $21.447 \pm 0.004$ \\
F814W  & $21.591 \pm 0.002$ & $21.624 \pm 0.001$ & $21.625 \pm 0.001$ & $21.598 \pm 0.001$ \\
F850LP & $19.899 \pm 0.005$ & $19.932 \pm 0.005$ & $19.934 \pm 0.005$ & $19.906 \pm 0.005$ \\
F1042M & $16.189 \pm 0.010$ & $16.222 \pm 0.010$ & $16.224 \pm 0.010$ & $16.196 \pm 0.010$ \\
\enddata
\end{deluxetable}

\clearpage

\begin{deluxetable}{lcccc}
\tabletypesize{\scriptsize}
\tablecaption{Zero Points for GAIN=14\label{tab_zp14}}
\tablewidth{0pt}
\tablehead{\colhead{Filter} & \colhead{PC} & \colhead{WFC2} & \colhead{WFC3} & \colhead{WFC4}}
\startdata
F380W  & $20.234 \pm 0.004$ & $20.255 \pm 0.003$ & $20.237 \pm 0.003$ & $20.258 \pm 0.003$ \\
F410M  & $18.595 \pm 0.004$ & $18.616 \pm 0.004$ & $18.598 \pm 0.004$ & $18.618 \pm 0.004$ \\
F439W  & $20.108 \pm 0.002$ & $20.129 \pm 0.001$ & $20.111 \pm 0.001$ & $20.132 \pm 0.002$ \\
F450W  & $21.204 \pm 0.004$ & $21.225 \pm 0.003$ & $21.207 \pm 0.003$ & $21.228 \pm 0.004$ \\
F467M  & $19.130 \pm 0.004$ & $19.151 \pm 0.004$ & $19.133 \pm 0.004$ & $19.153 \pm 0.004$ \\
F547M  & $20.867 \pm 0.003$ & $20.863 \pm 0.002$ & $20.871 \pm 0.002$ & $20.884 \pm 0.002$ \\
F555W  & $21.757 \pm 0.002$ & $21.753 \pm 0.001$ & $21.761 \pm 0.001$ & $21.774 \pm 0.002$ \\
F569W  & $21.430 \pm 0.003$ & $21.426 \pm 0.002$ & $21.434 \pm 0.002$ & $21.447 \pm 0.003$ \\
F606W  & $22.099 \pm 0.003$ & $22.095 \pm 0.002$ & $22.103 \pm 0.002$ & $22.116 \pm 0.002$ \\
F622W  & $21.564 \pm 0.006$ & $21.579 \pm 0.005$ & $21.576 \pm 0.005$ & $21.576 \pm 0.006$ \\
F675W  & $21.268 \pm 0.002$ & $21.282 \pm 0.001$ & $21.280 \pm 0.001$ & $21.279 \pm 0.003$ \\
F702W  & $21.667 \pm 0.004$ & $21.682 \pm 0.003$ & $21.679 \pm 0.003$ & $21.679 \pm 0.004$ \\
F785LP & $19.902 \pm 0.003$ & $19.915 \pm 0.003$ & $19.908 \pm 0.003$ & $19.912 \pm 0.003$ \\
F791W  & $20.707 \pm 0.004$ & $20.720 \pm 0.004$ & $20.713 \pm 0.004$ & $20.717 \pm 0.004$ \\
F814W  & $20.857 \pm 0.001$ & $20.870 \pm 0.001$ & $20.863 \pm 0.001$ & $20.867 \pm 0.001$ \\
F850LP & $19.166 \pm 0.005$ & $19.179 \pm 0.005$ & $19.172 \pm 0.005$ & $19.176 \pm 0.005$ \\
F1042M & $15.456 \pm 0.010$ & $15.468 \pm 0.010$ & $15.462 \pm 0.010$ & $15.466 \pm 0.010$ \\
\enddata
\end{deluxetable}

\clearpage

\begin{deluxetable}{lcccccccccccccc}
\rotate
\tabletypesize{\scriptsize}
\tablecaption{$BVRI$ Transformations\label{tab_xform}}
\tablewidth{0pt}
\tablecolumns{15}
\tablehead{
\colhead{} & \colhead{} & \colhead{} & \colhead{} & \colhead{} & \multicolumn{4}{c}{GAIN=7 Zero Points} & \multicolumn{4}{c}{GAIN=14 Zero Points} & \colhead{} & \colhead{} \\
\colhead{Filter} & \colhead{SMAG} & \colhead{SCOL} & \colhead{T1} & \colhead{T2} & \colhead{PC} & \colhead{WFC2} & \colhead{WFC3} & \colhead{WFC4} & \colhead{PC} & \colhead{WFC2} & \colhead{WFC3} & \colhead{WFC4} & \colhead{Cmin} & \colhead{Cmax}
}
\startdata
F380W  & B & (B-V) & -0.581 &  0.777 & 21.025 & 21.037 & 21.037 & 21.021 & 20.259 & 20.280 & 20.262 & 20.283 &      &  0.5 \\
F380W  & B & (B-V) & -0.943 &  0.103 & 21.377 & 21.389 & 21.389 & 21.373 & 20.611 & 20.632 & 20.614 & 20.635 &  0.5 &      \\
F410M  & B & (B-V) & -0.183 & -0.287 & 19.670 & 19.683 & 19.683 & 19.666 & 18.905 & 18.926 & 18.908 & 18.928 &      &      \\
F439W  & B & (U-B) & -0.103 & -0.046 & 20.861 & 20.873 & 20.873 & 20.857 & 20.095 & 20.116 & 20.098 & 20.119 &      &      \\
F439W  & B & (B-V) &  0.003 & -0.088 & 20.874 & 20.886 & 20.886 & 20.870 & 20.108 & 20.129 & 20.111 & 20.132 &      &      \\
F439W  & B & (B-R) &  0.019 & -0.049 & 20.868 & 20.880 & 20.880 & 20.864 & 20.102 & 20.123 & 20.105 & 20.126 &      &      \\
F439W  & B & (B-I) &  0.005 & -0.023 & 20.871 & 20.883 & 20.883 & 20.867 & 20.105 & 20.126 & 20.108 & 20.129 &      &      \\
F450W  & B & (B-V) &  0.230 & -0.003 & 21.972 & 21.984 & 21.984 & 21.968 & 21.206 & 21.227 & 21.209 & 21.230 &      &      \\
F467M  & B & (B-V) &  0.480 & -0.299 & 19.902 & 19.915 & 19.915 & 19.898 & 19.137 & 19.158 & 19.140 & 19.160 &      &  0.5 \\
F467M  & B & (B-V) &  0.432 & -0.002 & 19.853 & 19.866 & 19.866 & 19.849 & 19.088 & 19.109 & 19.091 & 19.111 &  0.5 &      \\
F547M  & V & (B-V) &  0.031 & -0.039 & 21.613 & 21.622 & 21.625 & 21.604 & 20.862 & 20.858 & 20.866 & 20.879 &      &  1.0 \\
F547M  & V & (B-V) &  0.056 & -0.016 & 21.565 & 21.574 & 21.577 & 21.556 & 20.814 & 20.810 & 20.818 & 20.831 &  1.0 &      \\
F547M  & V & (V-I) &  0.027 & -0.032 & 21.613 & 21.622 & 21.625 & 21.604 & 20.862 & 20.858 & 20.866 & 20.879 &      &  1.1 \\
F547M  & V & (V-I) &  0.049 & -0.013 & 21.565 & 21.574 & 21.577 & 21.556 & 20.814 & 20.810 & 20.818 & 20.831 &  1.1 &      \\
F555W  & V & (U-V) & -0.014 &  0.005 & 22.489 & 22.498 & 22.501 & 22.480 & 21.738 & 21.734 & 21.742 & 21.755 &      &      \\
F555W  & V & (B-V) & -0.060 &  0.033 & 22.508 & 22.517 & 22.520 & 22.499 & 21.757 & 21.753 & 21.761 & 21.774 &      &      \\
F555W  & V & (V-R) & -0.121 &  0.120 & 22.513 & 22.522 & 22.525 & 22.504 & 21.762 & 21.758 & 21.766 & 21.779 &      &      \\
F555W  & V & (V-I) & -0.052 &  0.027 & 22.508 & 22.517 & 22.520 & 22.499 & 21.757 & 21.753 & 21.761 & 21.774 &      &      \\
F569W  & V & (V-I) &  0.089 & -0.003 & 22.179 & 22.188 & 22.191 & 22.170 & 21.428 & 21.424 & 21.432 & 21.445 &      &  2.0 \\
F569W  & V & (V-I) & -0.125 &  0.022 & 22.511 & 22.520 & 22.523 & 22.502 & 21.760 & 21.756 & 21.764 & 21.777 &  2.0 &      \\
F606W  & V & (B-V) &  0.293 &  0.015 & 22.859 & 22.868 & 22.871 & 22.850 & 22.108 & 22.104 & 22.112 & 22.125 &      &  1.7 \\
F606W  & V & (B-V) & -0.284 &  0.081 & 23.649 & 23.658 & 23.661 & 23.640 & 22.898 & 22.894 & 22.902 & 22.915 &  1.7 &      \\
F606W  & V & (V-I) &  0.254 &  0.012 & 22.859 & 22.868 & 22.871 & 22.850 & 22.108 & 22.104 & 22.112 & 22.125 &      &  2.0 \\
F606W  & V & (V-I) & -0.247 &  0.065 & 23.649 & 23.658 & 23.661 & 23.640 & 22.898 & 22.894 & 22.902 & 22.915 &  2.0 &      \\
F622W  & R & (V-R) & -0.252 & -0.111 & 22.343 & 22.354 & 22.352 & 22.341 & 21.578 & 21.593 & 21.590 & 21.590 &      &      \\
F675W  & R & (U-R) &  0.039 & -0.007 & 22.053 & 22.064 & 22.062 & 22.050 & 21.288 & 21.302 & 21.300 & 21.299 &      &      \\
F675W  & R & (B-R) &  0.092 & -0.017 & 22.034 & 22.045 & 22.043 & 22.031 & 21.269 & 21.283 & 21.281 & 21.280 &      &      \\
F675W  & R & (V-R) &  0.253 & -0.125 & 22.033 & 22.044 & 22.042 & 22.030 & 21.268 & 21.282 & 21.280 & 21.279 &      &      \\
F675W  & R & (R-I) &  0.273 & -0.066 & 22.024 & 22.035 & 22.033 & 22.021 & 21.259 & 21.273 & 21.271 & 21.270 &      &      \\
F702W  & R & (V-R) &  0.343 & -0.177 & 22.437 & 22.448 & 22.447 & 22.435 & 21.672 & 21.687 & 21.684 & 21.684 &      &  0.6 \\
F702W  & R & (V-R) &  0.486 & -0.079 & 22.315 & 22.326 & 22.325 & 22.313 & 21.550 & 21.565 & 21.562 & 21.562 &  0.6 &      \\
F785LP & I & (V-I) &  0.091 &  0.020 & 20.638 & 20.671 & 20.673 & 20.645 & 19.905 & 19.918 & 19.911 & 19.915 &      &      \\
F791W  & I & (V-I) & -0.029 & -0.004 & 21.441 & 21.473 & 21.475 & 21.447 & 20.707 & 20.720 & 20.713 & 20.717 &      &  1.0 \\
F791W  & I & (V-I) & -0.084 &  0.011 & 21.482 & 21.514 & 21.516 & 21.488 & 20.748 & 20.761 & 20.754 & 20.758 &  1.0 &      \\
F814W  & I & (U-I) & -0.018 &  0.002 & 21.567 & 21.600 & 21.601 & 21.574 & 20.833 & 20.846 & 20.839 & 20.843 &      &      \\
F814W  & I & (B-I) & -0.031 &  0.007 & 21.587 & 21.620 & 21.621 & 21.594 & 20.853 & 20.866 & 20.859 & 20.863 &      &      \\
F814W  & I & (V-I) & -0.062 &  0.025 & 21.591 & 21.624 & 21.625 & 21.598 & 20.857 & 20.870 & 20.863 & 20.867 &      &      \\
F814W  & I & (R-I) & -0.112 &  0.084 & 21.591 & 21.624 & 21.625 & 21.598 & 20.857 & 20.870 & 20.863 & 20.867 &      &      \\
F850LP & I & (V-I) &  0.160 &  0.023 & 19.881 & 19.914 & 19.916 & 19.888 & 19.148 & 19.161 & 19.154 & 19.158 &      &      \\
F1042M & I & (V-I) &  0.350 &  0.022 & 16.136 & 16.169 & 16.171 & 16.143 & 15.403 & 15.415 & 15.409 & 15.413 &      &      \\
\enddata
\end{deluxetable}

\end{document}